\documentclass[fleqn,usenatbib]{mnras}

\usepackage{newtxtext,newtxmath}

\usepackage[T1]{fontenc}
\usepackage{ae,aecompl}

\usepackage{graphicx}	
\usepackage{amsmath}	
\usepackage{amssymb}	
\usepackage{amsfonts}
\usepackage{enumerate} 
\usepackage{colordvi} 
\usepackage{bm}
\usepackage{epsfig}
\usepackage{float}
\usepackage{verbatim}
\usepackage{subfig}
\usepackage[dvipsnames]{xcolor}
\usepackage{multirow,multicol}

\usepackage[colorinlistoftodos]{todonotes}
\usepackage{hyperref}

\newcommand{\bfk}{\mbox{\boldmath$k$}}
\newcommand{\bfp}{\mbox{\boldmath$p$}}
\newcommand{\bfq}{\mbox{\boldmath$q$}}

\defcitealias{Markovic:2019sva}{Paper~I}
\defcitealias{Bose:2019ywu}{Paper~III}

\title[Assessing non-linear models II]{Assessing non-linear models for galaxy clustering II: model validation and forecasts for Stage IV surveys}

\author[B. Bose et al.]{
Benjamin Bose$^{1}$\thanks{E-mail:benjamin.bose@unige.ch},
Alkistis Pourtsidou$^{2,3}$,
Katarina Markovi\v{c}$^{4,5}$,
Florian Beutler$^{4}$
\\
$^{1}$Departement de Physique Theorique, Universite de Geneve, 24 quai Ernest Ansermet, 1211 Geneve 4, Switzerland \\
$^{2}$School of Physics and Astronomy, Queen Mary University of London, Mile End Road, London E1 4NS, UK\\
$^{3}$Department of Physics \& Astronomy, University of the Western Cape, Cape Town 7535, South Africa\\
$^{4}$Institute of Cosmology \& Gravitation, University of Portsmouth,
Dennis Sciama Building, Burnaby Road, Portsmouth PO1 3FX, UK\\
$^{5}$Jet Propulsion Laboratory, California Institute of Technology, 4800 Oak Grove Drive, Pasadena, CA 91109, USA\\
}

\date{Accepted XXX. Received YYY; in original form ZZZ}

\pubyear{2019}

\begin{document}
\label{firstpage}
\pagerange{\pageref{firstpage}--\pageref{lastpage}}
\maketitle

\begin{abstract}
 Accurate modelling of  non-linear scales in galaxy clustering will be crucial for  data analysis of Stage IV galaxy surveys. A selection of competing non-linear models must  be made based on validation studies.  We provide a comprehensive set of forecasts of two different models for the halo redshift space power spectrum, namely the commonly applied TNS model and an effective field theory of large scale structure (EFTofLSS) inspired model. Using simulation data and a  least-$\chi^2$ analysis,  we determine ranges of validity for the models. We then conduct an exploratory Fisher analysis using the full anisotropic power spectrum to investigate parameter degeneracies. 
 We proceed to perform an MCMC analysis utilising the monopole, quadrupole, and hexadecapole spectra, with a restricted range of scales for the latter in order  to avoid biasing our growth rate, $f$, constraint. We find that the TNS model with a Lorentzian damping  and standard Eulerian perturbative modelling outperforms other variants of the TNS model. 
Our MCMC analysis finds that the EFTofLSS-based model may provide tighter marginalised constraints on  $f$ at $z=0.5$ and $z=1$ than the TNS model, despite having additional nuisance  parameters. However this depends on the range of scales used as well as  the fiducial values and priors on the EFT nuisance parameters. Finally, we extend previous work to provide a consistent comparison between the Fisher matrix and MCMC forecasts using the multipole expansion formalism, and find good agreement between them.
\end{abstract}



\section{Introduction}

The standard model of cosmology, $\Lambda$CDM, has been hugely successful in reproducing many cosmological observations such as the cosmic microwave background (CMB) \citep{Planck:2015xua} and the large scale structure of the universe (LSS) \citep{Anderson:2012sa,Song:2015oza,Beutler:2016arn}. The model relies on two fundamental theoretical assumptions: that general relativity holds on all physical scales and that the universe is homogeneous and isotropic on large scales. While $\Lambda$CDM fits observational data extremely well, it requires the introduction of two exotic dark components: cold dark matter (CDM) and dark energy in the form of a cosmological constant ($\Lambda$), which account for $95\%$ of the matter-energy content of the Universe today. Probing the nature of dark matter and dark energy is a key driver in modern cosmology, and a plethora of dark matter, exotic dark energy and modified gravity models have been proposed \citep[for respective reviews, see][]{Bertone:2004pz,Copeland:2006wr,Clifton:2011jh}.  
\newline
\newline
Large scale structure (LSS) measurements offer promising means of testing $\Lambda$CDM and gravity. In particular, the measurement of the  redshift space distortions (RSD) phenomenon in the galaxy distribution can put meaningful constraints on cosmology. This has traditionally been done by modeling the redshift space galaxy power spectrum or correlation function \citep{Blake:2011rj,Reid:2012sw,Macaulay:2013swa,Beutler:2013yhm,Gil-Marin:2015sqa,Simpson:2015yfa}. 
It is expected that very precise measurements of the observables will be made with the commencement of new, very large spectroscopic surveys such as EUCLID\footnote{\url{www.euclid-ec.org}} \citep{Blanchard:2019oqi}, WFIRST\footnote{\url{https://wfirst.gsfc.nasa.gov/}}, the Dark Energy Spectroscopic Instrument (DESI)\footnote{\url{www.desi.lbl.gov}} \citep{Aghamousa:2016zmz}, and the Square Kilometre Array (SKA)\footnote{\url{www.skatelescope.org/}} \citep{Bacon:2018dui}. 
\newline
\newline
In order to make the most of the upcoming data sets, theoretical models for the redshift space galaxy power spectrum must be studied carefully. Perturbation theory based models offer a robust and computationally quick means of modeling the RSD at large distance scales \citep{Bernardeau:2001qr,Kaiser:1987qv,Scoccimarro:2004tg}. Furthermore, they offer the flexibility to give predictions for a wide range of gravity and dark energy models \citep{Bose:2016qun,Bose:2017dtl,Bose:2017jjx,Bose:2018orj}. To extend their range of applicability, phenomenological ingredients can be added in order to model \emph{non-linear physics} \citep{Taruya:2010mx,Senatore:2014vja,delaBella:2017qjy}. 
Working in Fourier space and assuming a high degree of Gaussianity, the amount of information available in the matter power spectrum is roughly given by the number of independent modes we can access. 
Therefore, extending theoretical models to include non-linear scales should in principle allow us to extract much more information from data. However, this is heavily dependent on our ability to model non-linear structure formation in an unbiased way. 
\newline
\newline
On top of this, additional modeling is required to relate the dark matter and galaxy distributions, a relation called \emph{galaxy bias}. Such non-linear and galaxy bias modeling often come with so-called `nuisance' parameters, which are not known (up to some motivated priors) a priori. As their name implies, these parameters are not generally interesting and are marginalized over when constraining cosmology. This marginalization weakens our constraints, essentially leading us to an issue of optimization. We then must ask: \emph{What models give us an accurate description of the galaxy distribution over the largest range of scales but without invoking unnecessary degrees of freedom?} The issue of optimal power spectrum modeling has been recently studied in a number of works \citep{delaBella:2018fdb,Osato:2018ldv,Bose:2018orj} and will be the focus of this paper. 
\newline
\newline 
At the current forefront of perturbation theory based RSD modeling are two main approaches. The first is the so-called TNS model \citep{Taruya:2010mx}, which combined with the bias model of \citet{McDonald:2009dh} has been an integral part of the BOSS data analysis \citep{Beutler:2016arn}. This model has been studied extensively and has been shown to reproduce the broadband power spectrum including RSD from simulations at linear and moderately non-linear scales \citep{Nishimichi:2011jm,Taruya:2013my,Ishikawa:2013aea,Zheng:2016zxc,Gil-Marin:2015sqa,Gil-Marin:2015nqa,Bose:2017myh,Bose:2016qun,Markovic:2019sva}. 
\newline
\newline
The second is the effective field theory approach (EFT) commonly used in other fields of physics such as particle physics or condensed matter. The EFT of LSS (EFTofLSS) \citep{Baumann:2010tm,Carrasco:2012cv} represents an attempt to separate linear and non-linear physics so that one can safely model contributions from the small scale regime independently from the large scale contributions, as well as any back-reaction effects by the non-linear physics on the linear scales. The non-linear modeling comes with degrees of freedom in the form of sound speed parameters $c_{s}$. These parameters are time dependent coupling constants that arise from treating the stress energy tensor perturbatively and performing a time integral over the Green's function and associated kernels in order to get the corresponding contributions to the power spectrum. This approach has been shown to model simulation measurements down to much smaller scales than the standard perturbative approach \citep{Senatore:2014vja,Lewandowski:2015ziq,Perko:2016puo,Foreman:2015lca} and has become a promising means of modeling LSS. Recent bias models have also been developed within this framework \citep{Angulo:2015eqa,Perko:2016puo,Fujita:2016dne} but these generally come with many additional degrees of freedom. For example \citet{Perko:2016puo} models the RSD halo power spectrum with 10 nuisance parameters. 
\newline
\newline
In this work we consider a TNS-based model similar to that used in the BOSS survey \citep{Beutler:2016arn} and one of the EFTofLSS-based models used in \citet{delaBella:2018fdb}, but with a reduced nuisance parameter set. Using a set of COLA simulations \citep{Tassev:2013pn,Howlett:2015hfa, Valogiannis:2016ane,Winther:2017jof} we determine a range of validity for the models\footnote{ We have checked that the deviation of COLA from full N-body is sufficiently accurate for the scales of interest for the halo monopole we utilise in this work.}.  We then perform an exploratory Fisher matrix forecast analysis using the full anisotropic power spectrum $P(k,\mu)$ and specifications similar to forthcoming Stage IV spectroscopic surveys. The Fisher analysis allows the fast exploration of parameter space and the fast investigation of different assumptions. We focus on investigating parameter degeneracies and the effect of imposing priors on nuisance parameters, as well as providing estimates for the constraints we can expect on the \emph{logarithmic growth rate}, $f$. This parameter is strongly cosmology and gravity dependent, and represents the rate at which structure grows in a Friedman-Lemaitre-Robertson-Walker universe. We proceed to present various MCMC analyses which provide a more accurate test of parameter degeneracies and marginalised constraints. We finally follow previous studies \citep{Wolz:2012sr,Hawken:2011nd} and compare our EFTofLSS posterior probability distributions resulting from MCMC to that of the Fisher analysis.  We conduct the analysis using power spectrum multipoles, $P^{(S)}_l(k)$, in order to make it maximally comparable to real data analysis, as we
recommended in \citetalias{Markovic:2019sva} of this series \citep{Markovic:2019sva}.
\newline
\newline  
This paper is organized as follows: In \autoref{sec:models} we present the biased tracer RSD models. In \autoref{sec:sims-comparison} we present a comparison of model predictions with simulation data and determine fiducial nuisance parameters and a range of validity for each. In \autoref{sec:forecasts} we perform the exploratory Fisher analysis with our chosen models and present results, followed by the MCMC analysis and results in \autoref{sec:mcmc}. In \autoref{sec:FishMCMCcomp} we perform a comparison between Fisher matrix and MCMC forecasts using the multipole expansion formalism. In \autoref{sec:conclusions} we summarise our findings and conclude.


\section{Theoretical background and model selection}
\label{sec:models}

We begin by presenting the two models we will use in our forecasts. Both are based on standard Eulerian perturbation theory (SPT), which has the following core assumptions:
\begin{itemize}
\item 
We live on a spatially expanding, homogeneous and isotropic background spacetime. 
\item
We work on scales far within the horizon but at scales where $\delta, \theta \ll 1$, where $\delta$ and $\theta$ are the density and velocity perturbations respectively. This is the so called \emph{Newtonian regime} at quasi non-linear scales. 
\end{itemize}
In addition we assume that the gravitational interaction is described by general relativity \footnote{This assumption can be relaxed quite easily within SPT \citep[e.g][]{Bose:2016qun}.}. Aside from the above, each model includes phenomenological ingredients and a set of free parameters which will be made explicit in the following sections.

\subsection{TNS-based model}

The first is the TNS RSD model \citep{Taruya:2010mx} combined with the tracer bias model of \citet{McDonald:2009dh}. A similar model has been used in the BOSS analyses to infer cosmological constraints \citep{Beutler:2013yhm,Beutler:2016arn}, the exact differences from which will be made explicit soon. The model is given by 
 \begin{align}
 P^{S}_{\rm TNS}(k,\mu) =& D_{\rm FoG}(\mu^2 k^2 \sigma_v^2)\Big[ P_{g,\delta \delta} (k) \nonumber \\ & + 2 \mu^2 P_{g,\delta \theta}(k) +  \mu^4 P_{\theta \theta}^{\rm 1-loop} (k) \nonumber \\ & + b_1^3A(k,\mu) + b_1^4B(k,\mu) +b_1^2 C(k,\mu)  \Big], 
 \label{redshiftps}
 \end{align} 
where the superscript $S$ denotes the power spectrum in redshift space. The terms in brackets are all constructed within SPT, while the prefactor, $D_{\rm FoG}$, is added for phenomenological modeling of the fingers-of-god effect. Within this prefactor, $\sigma_v$ is a free parameter and represents the velocity dispersion of the cluster; $f$ is the logarithmic growth rate and $\mu$ is the cosine of the angle between $\bfk$ and the line of sight. The perturbative components of the model, along with the explicit dependency on the linear bias $b_1$, second order bias $b_2$ and constant stochasticity $N$ nuisance parameters, are given by \footnote{We make the local Lagrangian bias assumption \citep{Sheth:2012fc,Chan:2012jj,Saito:2014qha,Baldauf:2012hs}.} 
\begin{align}
P_{g,\delta \delta}(k) &=  b_1^2 P^{1-{\rm loop}}_{\delta \delta} (k) + D_1^4\Big[ 2 b_2 b_1 P_{b2,\delta}(k) \nonumber \\  & -\frac{8}{7} (b_1^2-b_1) P_{bs2,\delta}(k) + \frac{64}{315} (b_1^2-b_1)\sigma_3^2(k)P_L(k) \nonumber \\ & + b_2^2 P_{b22}(k) - \frac{8}{7} b_2 (b_1-1) P_{b2s2}(k) \nonumber \\  & + \frac{16}{49}(b_1-1)^2 P_{bs22} (k)\Big] + N, \label {pb1} \\
P_{g,\delta \theta}(k) &=  b_1 P^{1-{\rm loop}}_{\delta \theta} (k) + D_1^4\Big[2 b_2  P_{b2,\theta}(k) \nonumber \\  & -\frac{4}{7}(b_1-1) P_{bs2,\theta}(k) + \frac{32}{315}(b_1-1) \sigma_3^2(k)P_L(k) \Big], \label{pb2}
\end{align}
 where $D_1$ is the linear growth factor at the desired redshift $z$ and $P_L(k)$ is the primordial matter power spectrum. Note that there is no velocity bias, therefore $P_{g,\theta\theta}=P_{\theta \theta}^{\rm 1-loop}$.The 1-loop dark matter spectra are then given by
\begin{equation}
P^{1-{\rm loop}}_{ij}(k;a)  = F_{ij}\Big[ D_1^2P_L(k) + D_1^4P^{22}_{ij}(k) + D_1^4P^{13}_{ij}(k)\Big],
\label{eq:1loop}
\end{equation}
where $i,j \in \{\delta,\theta\}$ and $F_{\delta \delta} = 1, F_{\delta \theta} = f$ and $F_{\theta \theta}=f^2$. The components are further expanded in terms of the standard Einstein-de Sitter perturbative kernels $F_2,F_3,G_2$ and $G_3$ \citep{Bernardeau:2001qr} as 
\begin{align}
P_{\delta \delta}^{22}(k)
&= \frac{2}{(2\pi)^{3}}\int d^3 q F_2 (\bfk - \bfq, \bfq)^2 
P_L(|\bfk - \bfq|) P_L(q), \\
P_{\delta \theta}^{22}(k) 
&= \frac{2}{(2\pi)^{3}} \int d^3 q F_2 (\bfk - \bfq, \bfq) G_2 (\bfk - \bfq, \bfq) \nonumber \\ & \qquad \qquad  \times 
P_L(|\bfk - \bfq|) P_L(q), \\
P_{\theta \theta}^{22}(k) 
& = \frac{2}{(2\pi)^{3}} \int d^3 q G_2 (\bfk - \bfq, \bfq)^2
P_L(|\bfk - \bfq|) P_L(q),
\end{align}
and 
\begin{align}
P_{\delta \delta}^{13}(k) 
&= \frac{6}{(2\pi)^{3}} \int d^3 q F_3(\bfk, \bfq, -\bfq) 
P_L(q) P_L(k), \\
P_{\delta \theta}^{13}(k) 
&= \frac{3}{(2\pi)^{3}}  \int d^3 q G_3 (\bfk, \bfq, -\bfq)
P_L(q)  P_L(k) \nonumber \\  & \quad \quad+ 3  \int d^3 q F_3 (\bfk, \bfq, -\bfq) 
P_L(q) P_L(k), \\
P_{\theta \theta}^{13}(k) 
& = \frac{6}{(2\pi)^{3}} \int d^3 q G_3 (\bfk, \bfq, -\bfq) 
P_L(|\bfk - \bfq|) P_L(q).
\end{align}
The RSD correction terms, $A(k,\mu)$, $B(k,\mu)$ and $C(k,\mu)$ are given by
\begin{align}
A(k,\mu) =& \ D_1^4\sum_{m,n=1}^3 \mu^{2m} f^{n} \frac{ k^3}{(2\pi)^2} \nonumber \\ 
    &\times \bigg[ \int dr \int dx \Big( A_{mn}(r,x) P_L(k) + \tilde{A}_{mn}(r,x) P_L(kr)\Big)  \nonumber \\ 
    &\times \frac{P_L(k\sqrt{1+r^2-2rx})}{(1+r^2-2rx)} + P_L(k)\int dr a_{mn}(r)P_L(kr) \bigg], \\
B(k,\mu) =& \ D_1^4 \sum_{n=1}^4 \sum_{a,b=1}^2 \mu^{2n}(-f)^{a+b}\frac{k^3}{(2\pi)^2} \nonumber \\  
    &\times \int dr \int dx     B^n_{ab}(r,x)\frac{P_{a2}(k\sqrt{1+r^2-2rx}) P_{b2}(kr)}{(1+r^2-2rx)^a}, \\
C(k,\mu) =& \ D_1^4 (k\mu f)^2 \nonumber \\ 
    &\times \int \frac{d^3p d^3q}{(2\pi)^3}\delta_D(\bfk-\bfq-\bfp)\frac{\mu_p^2}{p^2}(1+fx^2)^2P_L(p)P_L(q),
\label{cterm}
\end{align}
where $\mu_p =\hat{\bfk}\cdot \hat{\bfp}$, $r=k/q$ and $x=\hat{\bfk}\cdot \hat{\bfq}$. Explicit expressions for $A_{mn},\tilde{A}_{mn},a_{mn}$ and $B_{ab}^n$ can be found in the Appendices of \citet{Taruya:2010mx}. The $C(k,\mu)$ term is known to have small enough acoustic features so it is usually omitted in the literature. It can be effectively absorbed into the fingers-of-god prefactor of \autoref{redshiftps}. In our analysis we include it. Finally, the bias terms are given by 
\begin{align}
P_{b2,\delta}(k) =& \int \frac{d^3 q}{(2\pi)^3} P_L(q) P_L(|\bfk-\bfq|) F_2(\bfq,\bfk-\bfq), \label{b1} \\
P_{b2,\theta}(k) =& \int \frac{d^3 q}{(2\pi)^3} P_L(q) P_L(|\bfk-\bfq|) G_2(\bfq,\bfk-\bfq), \label{b2} \\
P_{bs2,\delta}(k) =& \int \frac{d^3 q}{(2\pi)^3} P_L(q) P_L(|\bfk-\bfq|) F_2(\bfq,\bfk-\bfq) S^{(2)}(\bfq,\bfk-\bfq),\label{b3} \\
P_{bs2,\theta}(k) =& \int \frac{d^3 q}{(2\pi)^3} P_L(q) P_L(|\bfk-\bfq|) G_2(\bfq,\bfk-\bfq) S^{(2)}(\bfq,\bfk-\bfq), \label{b4}\\
P_{b22}(k) =& \frac{1}{2}\int \frac{d^3 q}{(2\pi)^3} P_L(q) \Big[P_L(|\bfk-\bfq|) -P_L(q) \Big],\label{b5} \\
P_{b2s2}(k) =& -\frac{1}{2}\int \frac{d^3 q}{(2\pi)^3} P_L(q) \nonumber \\ &  \quad \quad \times \Big[ \frac{2}{3}P_L(q) -  P_L(|\bfk-\bfq|)S^{(2)}(\bfq,\bfk-\bfq)  \Big], \label{b6}\\
P_{bs22}(k) =& -\frac{1}{2}\int  \frac{d^3 q}{(2\pi)^3} P_L(q) \nonumber \\ &  \quad \quad \times \Big[ \frac{4}{9}P_L(q) -  P_L(|\bfk-\bfq|)S^{(2)}(\bfq,\bfk-\bfq)^2  \Big],\label{b7} \\
\sigma_3^2(k) =& \frac{210}{112}\int\frac{d^3 q}{(2\pi)^3} P_L(q) \nonumber \\ &  \quad \quad \times  \Big[\Big(S^{(2)}(-\bfq,\bfk)-\frac{2}{3}\Big)S^{(2)}(\bfq,\bfk-\bfq) + \frac{4}{9}\Big], \label{b8}
\end{align}
where the additional kernel $S^{(2)}$ is given by 
\begin{equation}
S^{(2)}(\bfq_1,\bfq_2) = -\frac{1}{3}(1-3\mu_{1,2}^2),
\end{equation}
where $\mu_{1,2}$ is the cosine of the angle between $\bfq_1$ and $\bfq_2$. 
Since we only consider moderately non-linear scales and redshifts at or above $z=0.5$, where non-linearity is weak, the following assumptions we have made are valid: 
\begin{enumerate}
\item
Negligible velocity bias, i.e. $\theta_g = \theta_m$. 
\item
The local Lagrangian assumption \citep[as validated by N-body simulations,][]{Baldauf:2012hs}. This allows us to reduce the number of free bias parameters from 5 to 3.
\item
The Einstein-de Sitter approximation in the perturbative calculations allowing us to separate time and scale components of the perturbations. This is well known to be an excellent approximation for GR \citep{Bose:2016qun,Bose:2018orj}. 
\end{enumerate}
Furthermore, we will investigate two functional forms for the $D_{\rm FoG}$ term, a Lorentzian and a Gaussian:
\begin{align}
D_{\rm FoG}^{\rm Lor}(k^2\mu^2 \sigma_v^2) &= \frac{1}{1 + (k^2\mu^2 \sigma_v^2)/2} \, , \nonumber \\
D_{\rm FoG}^{\rm Gau}(k^2\mu^2 \sigma_v^2) &= \exp{[-(k^2\mu^2 \sigma_v^2)]} \, .
\end{align}
The key differences between this model and that used in the galaxy clustering data analysis of \citet{Beutler:2016arn} for example, is the inclusion of the $C(k,\mu)$ term and the fact that we use SPT instead of the RegPT prescription of \citet{Taruya:2012ut} for the 1-loop dark matter power spectra (\autoref{eq:1loop}). In that analysis they choose the Gaussian form for $D_{\rm FoG}$. Furthermore, the TNS model is similar to the \texttt{M\&R+SPT} model considered in \citet{delaBella:2018fdb}. In that model they only consider the Gaussian damping factor shown above and do not assume the local Lagrangian picture. Further, they exclude $N$, giving their bias model $4$ degrees of freedom. We choose instead to use the bias model as used in the BOSS analysis in \citet{Beutler:2016arn}. 
\\ \\
The full set of {\bf nuisance parameters} in the TNS-based model we use is therefore $\{ \sigma_v, b_1, b_2, N \}$.

\subsection{EFTofLSS-based Model}

The second model we consider is based on the EFTofLSS prescription for the redshift space dark matter spectrum \citep{delaBella:2017qjy} given by 
\begin{align}
P^{S}_{\rm EFT}(k,\mu) =
    &\ P^{S}_{\rm SPT}(k,\mu) -\frac{2k^2}{k_{\rm NL}^2}D_1^2P_L(k) \nonumber \\ 
    &\times \bigg[c^2_{s,0} + c^2_{s,2} \mu^2 + c^2_{s,4} \mu^4 + \nonumber \\ 
    &+  \mu^6 \left(f^3 c^2_{s,0} - f^2 c^2_{s,2} + f c^2_{s,4}\right)\bigg],
\end{align}
where $c^2_{s,i}$ are the sound speed parameters of EFTofLSS and $k_{NL}^2$ indicates the strong coupling scale. None of these can be calculated, so they are usually measured as the combination $c^2_{s,i}/k^2_{NL}$. The $P^{S}_{SPT}(k,\mu)$ is the 1-loop SPT prediction for the redshift space power spectrum. As in \citet{delaBella:2017qjy}, a resummation technique \citep{Vlah:2015zda} is applied to the 1-loop spectra. The $P^S_{\rm SPT}$ is almost identical to \autoref{redshiftps} with $b_1=1$, $b_2 = N = 0$ and the phenomenological exponential prefactor now given by the SPT prediction $\left[1-\left(D_1^2f^2k^2\mu^2 \tilde{\sigma}^2_{v}\right)\right]$, where 
\begin{equation}
\tilde{\sigma}^2_{v} = \frac{1}{6\pi^2} \int dq P_L(q) \ .
\end{equation}
Also note that this prefactor does not multiply the correction terms $A,B$, and $C$ (see \autoref{redshiftps}).
\\
\\
The power spectrum model suggested here simply upgrades the redshift space dark matter spectrum $P^{S}_{\rm SPT}(k,\mu)$ to a biased tracer spectrum by using the bias model of \citet{McDonald:2009dh}. In this way we are only really adding EFTofLSS-like counter terms (terms involving $c^2_{s,i}$) to the SPT predicted redshift space halo spectrum. This model is very similar to the \texttt{EFT+M\&R} model considered in \citet{delaBella:2018fdb} with the difference that we omit the stochastic EFTofLSS terms that introduce an additional 3 nuisance parameters.
The explicit expression is  
\begin{align}
P^{S}_{\rm EFT}(k,\mu) =& \ \left\{1-\left(D_1^2f^2k^2\mu^2 \tilde{\sigma}^2_{v}\right)\right\} \nonumber \\
&\times \left[ P_{g,\delta \delta} (k) + 2  \mu^2 P_{g,\delta \theta}(k) +  \mu^4 P^{1-{\rm loop}}_{\theta \theta} (k) \right]  \nonumber \\ 
&+ b_1^3A(k,\mu) + b_1^4B(k,\mu) + b_1^2C(k,\mu) \nonumber \\ 
&- 2D_1^2P_L(k)k^2\Big[c^2_{s,0} + {c}^2_{s,2} \mu^2 + {c}^2_{s,4} \mu^4 \nonumber \\ 
&+ \mu^6 \left(f^3 {c}^2_{s,0} - f^2 {c}^2_{s,2} + f {c}^2_{s,4}\right)\Big],
\label{redshiftps2}
\end{align}
where we have absorbed the $k_{NL}^2$ into the $c_{s,i}^2$. We can motivate \autoref{redshiftps2} by arguing that the bias is well described by the  \citet{McDonald:2009dh} model and  we are just missing a suppression of power coming from small cosmological scales that can be described by the dark matter EFTofLSS counter-terms. Before proceeding we make two comments on the model proposed here.
\newline 
\newline
 First, we have omitted the 3 stochasticity terms of the EFTofLSS redshift space spectrum (\citet{Baumann:2010tm, Lewandowski:2015ziq}). For the dark matter power spectrum, these terms go as $\sim k^4$ and hence are not expected to impact the predictions at the scales considered here. This was also investigated and confirmed in the analysis of \citet{delaBella:2017qjy}. For halos, the omission of the stochastic terms may have an effect on the fits, but as we are making the assumption that all bias physics is captured by the \citet{McDonald:2009dh} model, we do not consider them. Further, the most that these terms can improve the range of validity of the model is up to the regime where the 2-loop contributions become important \citep{Carrasco:2013sva}. This extension in scale is not expected to compensate the degradation of marginalised constraints from the inclusion of 3 additional parameters -- this could be checked but is not the focus of this work. For a complete treatment of bias within the EFTofLSS we direct the reader to \citet{Perko:2016puo}. This treatment comes with $10$ free parameters and given the Bayesian information criterion used in \citet{delaBella:2018fdb} it is unlikely to be favoured against a similar model with fewer free parameters. 
\\ \\ 
 Second, we have not performed a full infra-red resummation of the baryon acoustic oscillation features, but rather have only applied resummation to the 1-loop power spectra pieces in \autoref{redshiftps2}. Since we have checked that there are large biases in the prediction for $f$ incurred by increasing $k_{\rm max}$ beyond the determined values, we do not expect inaccuracies in the resummation applied to improve this validity range for the model. Further, in \cite{delaBella:2018fdb}, the authors find that the exclusion of resummation in a model very similar to that proposed here does not affect their fits to simulations.
\\ \\ 
The full set of {\bf nuisance parameters} in the EFTofLSS-based model we use is therefore $\{ b_1,b_2,N, {c}_{s,0}^2,{c}_{s,2}^2,{c}_{s,4}^2 \}$. This is an additional 2 parameters compared to the TNS approach described by \autoref{redshiftps}.
\newline
\newline 
In \autoref{redshiftps} and \autoref{redshiftps2} we can immediately see the dependency of the power spectrum on the model parameters. The logarithmic growth rate, $f$ is also explicit. Cosmological parameter dependence enters through the primordial power spectrum $P_L(k)$ with $\sigma_8$\footnote{$\sigma_8$ governs the amplitude of density perturbations at $8$ Mpc/$h$.} being completely degenerate with $D_1$. For our analysis in the next section, since we are focused on comparing the power spectrum models, we assume a $\Lambda$CDM expansion and fix cosmology as well as $D_1$ and $f$ to their known values. 

\section{Comparison to simulations}
\label{sec:sims-comparison}

\begin{figure*}
\centering
  \includegraphics[width=15cm,height=8cm]{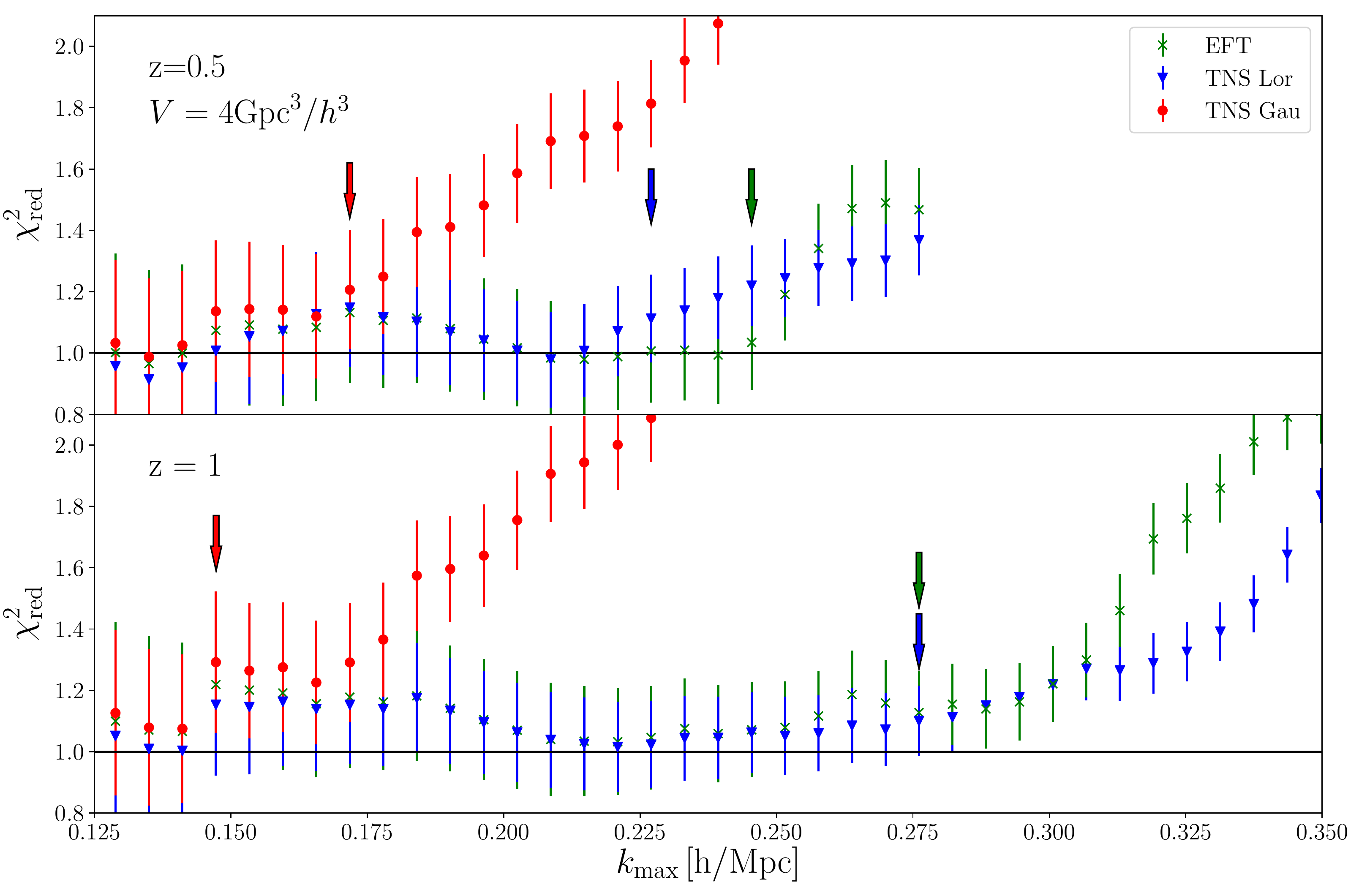}\\
  \caption[CONVERGENCE]{The minimized $\chi_{\rm red}^2$ statistic as a function of $k_{\rm max}$ at $z=0.5$ (top) and $z=1$ (bottom) for the EFTofLSS (green cross) and TNS model with a Lorentzian (blue triangle) and a Gaussian (red circle) $D_{\rm FoG}$ term. The error bars shown are the $2\sigma$ confidence interval for the $\chi_{\rm red}^2$ statistic with $N_{\rm dof}$ degrees of freedom. The arrows indicate the $k_{\rm max}$ value we use in our fits.}
\label{redc}
\end{figure*}

\begin{figure*}
\centering
  \includegraphics[width=15cm,height=8cm]{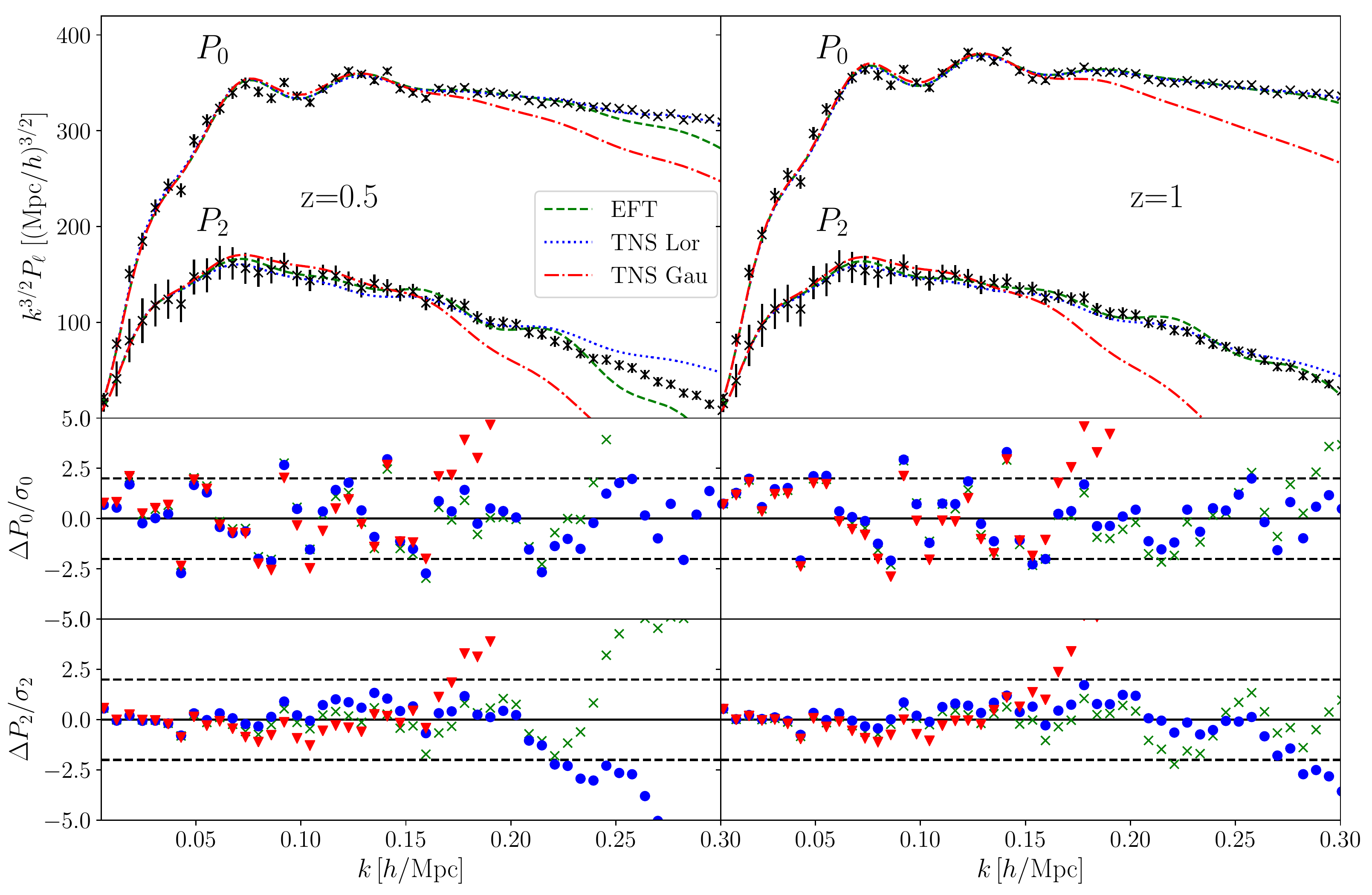} 
  \caption[CONVERGENCE]{The averaged halo monopole and quadrupole of 4 PICOLA simulations (black points) with errors given by linear theory assuming a survey volume of $V=4$\, Gpc${}^3/h^3$ and number density of $n_h = 1\times 10^{-3} \, h^3/\mbox{Mpc}^3$. The best fitting EFTofLSS (green dashed line) and TNS model with a Lorentzian (blue dotted line) and a Gaussian (red dot-dashed line) $D_{\rm FoG}$ are also shown. The lower panels show the monopole (middle) and quadrupole (bottom) residuals with the data of all 3 models, with markers and colours as in \autoref{redc}. The dashed lines indicate the $2\sigma$ region around the data.}
\label{fits}
\end{figure*}

\begin{figure}
\centering
  \includegraphics[scale=0.5]{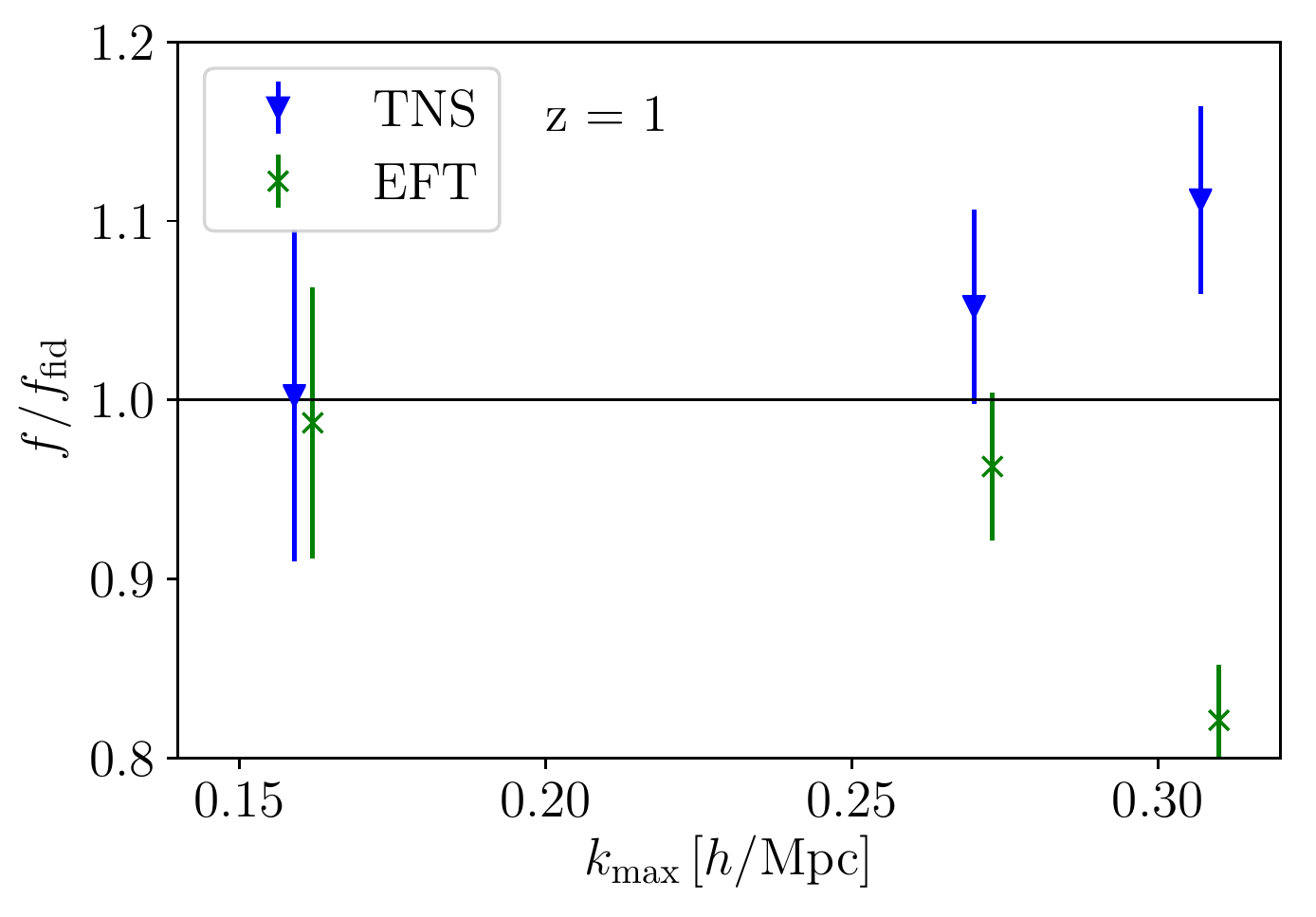} 
  \caption[]{Redshift space halo results at z = 1. The mean value of $f/f_{\rm fiducial}$ as a function of $k_{\rm max}$ using the
TNS (blue triangles) and EFTofLSS (green crosses) models with the marginalised $2\sigma$ error bars. Only $P_0$ and $P_2$ were
used in the analyses. Note that the EFTofLSS points have been slightly offset for better visualisation. 
}
\label{margplot}
\end{figure}

In this section we determine fiducial values for the nuisance parameters of each model described in the previous section as well as their respective ranges of validity. This is done by comparing to a set of Parallel COmoving Lagrangian Acceleration (PICOLA) simulations \citep{Howlett:2015hfa,Winther:2017jof}. Specifically, we use a set of four $\Lambda$CDM simulations of box length $1024 \, \mbox{Mpc}/h$ with $1024^3$ dark matter particles and a starting redshift $z_{\rm ini}=49$. The summed volume of these realisations is similar to Stage IV surveys such as DESI and Euclid at $z=1$ for a bin width of $\Delta z \sim 0.1$ \citep{Aghamousa:2016zmz,Majerotto:2012mf}.
\\ \\ 
The background cosmology in these DM-only simulations is taken from WMAP9 \citep{Hinshaw:2012aka}: $\Omega_m = 0.281$, $\Omega_b=0.046$, $h=0.697$, and $n_s=0.971$ and $\sigma_8(z=0) = 0.844$. We use halo catalogs, which were constructed using the friends-of-friends algorithm with a linking length of 0.2-times the mean particle separation. We consider the redshifts $z=0.5$ and $z=1$. For our analysis we use all halos above a mass of $M_{\rm min} = 4 \times 10^{12} \, M_{\odot}$. We note that the mass cut choice will affect the fiducial values and range of validity, and so we base our choice on the corresponding number density of halos at this mass cut which is $n_{\rm h} = 1\times 10^{-3} \, h^3/\mbox{Mpc}^3$. This number density is similar to that estimated for Stage IV surveys galaxy number density around the redshifts considered. 
\newline
\newline 
To determine the  fiducial values of the parameters and ranges of validity for the models we perform a fit to the simulated data using the redshift space power spectrum multipoles. PICOLA multipoles are measured using the distant-observer approximation\footnote{That is, we assume the observer is located at a distance much greater then the box size ($r\gg 1024 \, \mbox{Mpc}/h$), so we treat all the lines of sight as parallel to the chosen Cartesian axes of the simulation box. Next, we use an appropriate velocity component ($v_x, v_y$ or $v_z$) to disturb the position of a matter particle.} and averaged over three line-of-sight directions. We further average over the four PICOLA simulations.
\newline
\newline
On the theoretical side, the multipoles are defined as 
\begin{equation}
P_\ell^{(S)}(k)=\frac{2\ell+1}{2}\int^1_{-1}d\mu P^{S}(k,\mu)\mathcal{P}_\ell(\mu),
\label{eq:ell}
\end{equation}
where $\mathcal{P}_\ell(\mu)$ denote the Legendre polynomials and $P^{S}(k,\mu)$ is given by \autoref{redshiftps} or \autoref{redshiftps2}. For our fitting analysis, we utilise only the monopole ($\ell=0$) and quadrupole ($\ell=2$). The inclusion of the hexadecapole would significantly restrict the determined range in scale of validity and consequently the information gained since the monopole and quadrupole contain most of the RSD information. It is later considered in \autoref{sec:mcmc}, where we perform an MCMC analysis on the PICOLA data. 
\newline
\newline
To determine the range of validity, $k_{\rm max}$, that will be used to determine the fiducial parameters for each model, we follow the procedure outlined below:    
\begin{enumerate}
    \item
    We fix all cosmological parameters including the growth rate $f$ and perform a least-squares fit to the PICOLA data by varying the model nuisance parameters. We do this for all data bins within $0.125 \, h/{\rm Mpc} \leq k_{\rm max} \leq 0.300 \, h/{\rm Mpc}$.
    \item 
    We take the $95\%$ ($2\sigma$) confidence intervals ($2 \Delta \chi^2_{\rm red}$) on a $\chi^2$ distribution with $N_{\rm dof}$ degrees of freedom. Since $N_{\rm dof}$ is large in our analysis the errors are approximately symmetric. 
    \item 
  We determine $k_{\rm max}$ as the maximum k-value which has $[\chi^2_{\rm red}(k_{\rm max}) - 2\Delta \chi^2_{\rm red}(k_{\rm max})] \leq 1$.
\end{enumerate}
This gives a fair indication of the point at which the model gives a good fit to the data without biasing cosmology estimates\footnote{We test this by performing an MCMC analysis with $f$ being allowed to vary at the $k_{\rm max}$ determined here. For the TNS case, see \citetalias{Markovic:2019sva}.}. The reduced $\chi^2$ statistic is given by 
\begin{align}
\chi^2_{\rm red}(k_{\rm max}) & = \frac{1}{N_{\rm dof}}\sum_{k=k_{\rm min}}^{k_{\rm max}} \sum_{\ell,\ell'=0,2} \left[P^{S}_{\ell,{\rm data}}(k)-P^{S}_{\ell,{\rm model}}(k)\right]\nonumber \\ & \times \mbox{Cov}^{-1}_{\ell,\ell'}(k)\left[P^{S}_{\ell',{\rm data}}(k)-P^{S}_{\ell',{\rm model}}(k)\right],
\label{covarianceeqn}
\end{align}
 where $\mbox{Cov}_{\ell,\ell'}$ is the Gaussian covariance matrix between the different multipoles and $k_{\rm min}=0.006 \, h/{\rm Mpc}$. The number of degrees of freedom $N_{\rm dof}$ is given by $N_{\rm dof} = 2\times N_{\rm bins} - N_{\rm params}$, where $N_{\rm bins}$ is the number of $k-$bins used in the summation and $N_{\rm params}$ is the number of free parameters in the theoretical model. Here, $N_{\rm params} = 4$ for the TNS model of \autoref{redshiftps}, and $N_{\rm params} = 5$ for the EFTofLSS model of \autoref{redshiftps2}. The $N_{\rm params}$ is not $6$ for the EFTofLSS model because we only consider the monopole and quadrupole. When integrating to get each of these two multipoles, they come with the same k-dependent piece, $k^2 P(k)$, multiplied by a different linear combination of ${c}_{s,0}^2,{c}_{s,2}^2,{c}_{s,4}^2$. Therefore, by fixing any of the $c_{s,i}^2$, this constant can still take any value for each of $P_0$ and $P_2$ since the remaining two $c_{s,i}^2$ are still free to vary. Thus, one can have 3 independent fits for the first 3 multipoles using the EFTofLSS.  Finally, the bin-width we use is $\Delta k = 0.006 \, h/{\rm Mpc}$. 
\\ \\ 
We apply linear theory to model the covariance between the multipoles (see Appendix C of \citet{Taruya:2010mx} for details). This has been shown to reproduce N-body results up to $k\leq 0.300h/\mbox{Mpc}$ at $z=1$ (\cite{Taruya:2010mx})  and recently shown to work well at $z=0.5$ up to $k\leq 0.2h/{\rm Mpc}$ \cite{Sugiyama:2019ike}. In the covariance matrix we assume a number density of $n= 1\times 10^{-3} \, h^3/\mbox{Mpc}^3$ and a survey volume of $V_s=4 \, \mbox{Gpc}^3/h^3$. 
\newline
\newline
In \autoref{redc} we show the minimized $\chi^2_{\rm red}(k_{\rm max})$ for $z=0.5$ and $z=1$ for all the models considered, with their associated $2\sigma$ error bars. At both redshifts the Gaussian TNS model does significantly worse than the other two models with a rapidly increasing $\chi^2_{\rm red}$ for $k_{\rm max}>0.140 \, h/$Mpc. This was first studied in \citet{Sheth:1995is} and is not a new result. The other two models, EFTofLSS and TNS with a Lorentzian $D_{\rm FoG}$ do comparably well at $z=1$. This is expected as we have less non-linear structure formation at this time and so the additional parameters of the EFTofLSS model are not fully utilized. At $z=0.5$ on the other hand we find the EFTofLSS model does noticeably better than both TNS models. We show the $k_{\rm max}$ we choose for each model and the respective best fit parameters in \autoref{fittable}. These best fit models are plotted against the PICOLA data in \autoref{fits}. 
\newline 
\newline
We have also performed MCMC analyses to verify the $k_{\rm max}$ determined here at $z=1$. This is shown in \autoref{margplot}. Indeed, the models recover the simulation's fiducial value of $f$ at $k_{\rm max} = 0.276 \, h/{\rm Mpc}$ within 2$\sigma$. At $k_{\rm max}=0.305 \, h/{\rm Mpc}$ the TNS and EFTofLSS-like models become biased by over $5\sigma$ and $10\sigma$, respectively. This supports our determined $k_{\rm max}$ as being the scale at which the models truly start to break down. We also note that another restriction we could in principle impose would be to also recover the true value of the $b_1$ parameter. That means that the $\chi^2$ analysis could also be performed by fixing $b_1$ to that measured from the simulations. This could further constrain the models to scales where the bias model remains valid. We choose not to do this in our $\chi^2$ analysis, making our determined $k_{\rm max}$ optimistic -- but we will also perform Fisher matrix and MCMC analyses with more conservative $k_{\rm max}$ choices. For a similar analysis at $z=0.5$ and the study of recovering the value of the linear bias $b_1$, we refer the reader to Paper III of this series, \cite{Bose:2019ywu}. 
\newline
\newline
We have checked\footnote{This is calculated only using $P_0$ and $P_2$.} the $\chi^2_{\rm red}$ for the TNS model used in the BOSS analysis of \citet{Beutler:2016arn} up to the $k_{\rm max}$ we found in \autoref{fittable}. We remind the reader that this is different than the TNS model of \autoref{redshiftps}, as the $C(k,\mu)$ term is omitted and the 1-loop spectra are modeled with a RegPT prescription \citep{Taruya:2012ut}. Using this model, we find that at $z=1$:  
\begin{align}
\chi^2_{\rm red, Lor}(k_{\rm max}=0.276) &= 4.34 \pm (0.12) \ \rm{and} \\ \nonumber
\chi^2_{\rm red, Gau}(k_{\rm max}=0.147) &= 1.73 \pm (0.23) \ ,\nonumber
\end{align}
while at $z=0.5$ we find
\begin{align}
\chi^2_{\rm red, Lor}(k_{\rm max}=0.227) &= 1.37 \pm (0.14) \ \rm{and} \\ \nonumber
\chi^2_{\rm red, Gau}(k_{\rm max}=0.172) &= 1.39 \pm (0.19) \ , \nonumber
\end{align}
with the quoted errors being $2\sigma$, taken from the $\chi^2_{\rm red}$ distribution\footnote{At these $k_{\rm max}$ the SPT-TNS models considered here, \autoref{redshiftps}, have $\chi^2_{\rm red} = 1$ within $2\sigma$.}. It is therefore evident that the RegPT without $C(k,\mu)$ model does significantly worse in fitting the data at $z=1$ than the SPT based model, and marginally worse at $z=0.5$. We have checked that the $C(k,\mu)$ term does not affect the fit significantly which indicates a RegPT prescription in the TNS model is not optimal at redshifts $z\geq 0.5$. We should also point out that in the BOSS analysis the hexadecapole was included and it is undetermined if this would affect the relative RegPT and SPT best fits.
\newline
\newline
The RegPT prescription as used in the BOSS analysis of \citet{Beutler:2016arn} offers a damping of the 1-loop spectra once non-linearities become important, a feature that helps to avoid well known divergences in the SPT prescription \citep{Carlson:2009it,Nishimichi:2008ry} at low $z$. Our results suggest that the RegPT damping actually worsens the fit at redshifts where the SPT divergences are under control. This could also be partly because of the $D_{\rm FoG}$ factor which already provides small scale damping. For more details we refer to Appendix A of \citet{Bose:2017dtl} where we can clearly see the velocity spectra of SPT doing better than those of RegPT at $z=0.5$. We can also see SPT doing better at $z=1$ in Figure 2 of \citet{Osato:2018ldv}. It is worth noting though that adding an additional, phenomenological free damping parameter (similar to what is done for the TNS model), as in \citet{Osato:2018ldv}, a RegPT prescription can do better in modelling the small scales than EFTofLSS, RegPT and SPT, with respect to the matter power spectrum in real space. This is expected as we have introduced an additional degree of freedom by doing this. 
\begin{table}
\centering
\caption{Number of bins, $k_{\rm max} [h/{\rm Mpc}]$ and fiducial parameters for TNS and EFTofLSS models found by a least $\chi^2$ fit to the PICOLA data.}
\begin{tabular}{| c || c | c || c | c || c | c |}
\hline  
 \multicolumn{1}{ | c || }{Model} & \multicolumn{2}{|c||}{TNS Lor} & \multicolumn{2}{|c||}{TNS Gau} &  \multicolumn{2}{|c|}{EFTofLSS} \\
 \hline
 z & $0.5$ & $ 1 $ & $0.5$ & $1$ & $0.5$ & $1$ \\ \hline\hline 
 $N_{\rm bins}$ & $36$ & $44$ & $27$ & $23$  & $40$ & $44$  \\ \hline 
 $k_{\rm max}$ & $0.227$ & $0.276$ & $0.172$ & $0.147$  & $0.245$ & $0.276$  \\ \hline 
 
 $b_1$ &  $1.506$ & $1.897$ & $1.464$ & $1.918$  & $1.471$ & $1.905$   \\ \hline 
 
 $b_2$ & $0.091$ & $-0.318$ & $-0.741$ & $0.347$  & $-0.393$ & $-0.472$ \\ \hline 
 
 $N$ & $-272$ & $504$ & $847$ & $-60$ & $676$ & $274$ \\ \hline 
 $\sigma_v$ & $8.99$ & $8.09$ & $5.29$ & $5.17$  & - & -   \\ \hline 
  ${c}_{s,0}^2$ & - & - & - & -  & $2.718$ & $10^{-5}$  \\ \hline 
 ${c}_{s,2}^2$ & - & - & - & -  & $23.218$ & $3.121$  \\ \hline 
 ${c}_{s,4}^2$ & - & - & - & -  & $19.540$ & $42.750$ \\  
\hline
\end{tabular}
\label{fittable}
\end{table}

\noindent
\newline
\newline
\newline
Before moving forward we give some details on the $\chi^2_{\rm red}$ fits procedure:
\begin{enumerate}
    \item
    We perform initial fits using {\tt Mathematica}'s {\tt Minimize} function at a $k_{\rm max}= 0.125 \, h/{\rm Mpc}$.
    \item
    Using these best fit parameter values we perform a fast and crude search for better fits using the {\tt c++} code {\tt MG-COPTER} presented in \citet{Bose:2016qun}. This involves running $400,000$ $\chi_{\rm red}^2$ computations and accepting values with a lower $\chi^2_{\rm red}$ than the previous one. The least $\chi^2_{\rm red}$ of the run is stored\footnote{The step size in these searches is set to be reasonably large and is halved after half of the computations have been completed to improve efficiency.}. 
    \item  
    We run 5 additional searches with varying initial nuisance parameter values and check that they converge to the same value as the initial search\footnote{This is the case for most searches, but sometimes the additional searches achieve a slightly lower $\chi^2_{\rm red}$ than the initial search. In this case we use this lower value.}.
    \item  
    Using the best fit parameter values found above, steps 2 and 3 are repeated for a slightly larger $k_{\rm max}$ until all data bin values in $0.125 \, h/{\rm Mpc} \leq k_{\rm max}\leq 0.300 \,h/{\rm Mpc}$ are used. All steps are repeated for both redshifts. 
\end{enumerate}
 We also impose a flat positivity prior on the parameters: $ b_1, \sigma_v, {c}_{s,i} \geq 0 $. The results of this procedure are shown in \autoref{redc}. We should note that the method used here to determine $k_{\rm max}$ is fast but not ideal as we do not vary $f$, which has significant degeneracies with some nuisance parameters (see following sections). Also, the error bars we employ, taken from a $\chi^2$ distribution, are not very realistic. Ideally we would want to vary $f$ in a full MCMC analysis and then determine when the model recovers biased estimates. This is done in \citetalias{Bose:2019ywu} of this series \citep{Bose:2019ywu}, where the authors investigate the model's performance, parameter degenerecies, and marginalised $f$ constraints as a function of $k_{\rm max}$ by performing a large number of MCMC analyses on another set of PICOLA simulations. 
  

\section{Exploratory Fisher matrix forecasts}
\label{sec:forecasts}

In this section we are going to present forecasted constraints on the structure growth, $f$, in the TNS and EFTofLSS-based models presented previously, using the Fisher matrix formalism for the 2D anisotropic redshift space power spectrum $P(k,\mu)$. We do this, since it is an informative way to quickly gain an understanding of the correlations in a high-dimensional parameter space, as well as to conduct an exploratory analysis of the optimal setup of the problem. After we perform our MCMC analysis using the monopole, quadrupole, and hexadecapole spectra, we will perform another Fisher matrix analysis, this time using the multipole expansion formalism, which has been shown to be much more appropriate for comparison to real data analysis \citepalias{Markovic:2019sva}.
We begin by briefly describing the formalism, and then we move on to present our results. We note that  Fisher matrix codes used in this work are available from \url{https://github.com/Alkistis/GC-Fish-nonlinear}.

\subsection{Fisher matrix formalism for $P(k,\mu)$}

The Fisher matrix for a set of parameters $\{p\}$ is given by \citep{Fisher:1935,Tegmark:1997rp, Seo:2007ns}
\begin{equation}
F_{\rm ij} = \frac{1}{2}\left[C^{-1}\frac{\partial C}{\partial p_i}C^{-1}\frac{\partial C}{\partial p_j}\right] 
+ \frac{\partial \mathcal{M}^T}{\partial p_i}C^{-1}\frac{\partial \mathcal{M}^T}{\partial p_j}    \, ,    
\end{equation}
where $C$ is the covariance matrix and $\mathcal{M}$ the model of our observable.
The minimum errors on parameter $p_i$, marginalised over all other parameters, are given by the square root of the diagonal of the inverse of the Fisher matrix as
\begin{equation}
\Delta p_i \geq \sqrt{(F^{-1})_{ii}} \, .    
\end{equation}
This is known as the Cramer-Rao inequality: the diagonal elements of the inverse of the Fisher matrix give the best possible constraints we can achieve. Note that these are fully marginalised errors, including correlations with all other parameters. The unmarginalised ones are simply given by $\Delta p_i = 1/\sqrt{F_{ii}}$. Here we will focus on the full marginalised errors on $f$, the cosmological parameter of interest.
\\ \\
Following \citet{Feldman:1993ky}, we can write 
$\mathcal{M}_n \approx P^{\rm S}(k_n)$ in a thin Fourier shell of radius $k_n$, with $P^{\rm S}$ being the power spectrum signal. We can also write
\begin{equation}
C_{mn} \approx \frac{2}{V_nV_{\rm s}}[P^{\rm S}(k_n)+1/n]^2\delta_{\rm mn} \, ,
\end{equation} where $V_n \equiv 4\pi k^2_n dk_n/(2\pi)^3$ is the volume element and $dk_n$ the width of the shell. For convenience we can define the ``effective volume'' as
\begin{equation}
V_{\rm eff}(k_n) \equiv \left[\frac{nP^{\rm S}(k_n)}{1+nP^{\rm S}(k_n)}\right]^2 V_{\rm s} \, , \end{equation} 
with $n$ the number density of galaxies and $V_{\rm s}$ the survey volume.
For thick shells that contain many uncorrelated modes the Fisher Matrix can be written as \citep{Tegmark:1997rp}
\begin{equation}
F_{\rm ij} \approx \frac{1}{4\pi^2}\int^{k_{\rm max}}_{k_{\rm min}} k^2dk \; 
\frac{\partial {\rm ln}P^{\rm S}} {\partial p_i} 
\frac{\partial {\rm ln}P^{\rm S}} {\partial p_j}
V_{\rm eff} \, .
\label{eq:Fish1}
\end{equation} 
Considering the full power spectrum signal in redshift space, the Fisher matrix becomes \citep{Tegmark:1997rp, Seo:2007ns}
\begin{equation}
F_{\rm ij}=\frac{1}{8\pi^2}\int^{1}_{-1} d\mu \int^{k_{\rm max}}_{k_{\rm min}} k^2dk \; 
\frac{\partial {\rm ln}P^{\rm S}} {\partial p_i} 
\frac{\partial {\rm ln}P^{\rm S}} {\partial p_j}
V_{\rm eff} \, .
\label{eq:Fish2}    
\end{equation}
A useful quantity that we are going to utilise to present results is the \emph{correlation coefficient} $r$ given by
\begin{equation}
r(p_i,p_j) = \frac{(F^{-1})_{ij}}{\sqrt{(F^{-1})_{ii}(F^{-1})_{jj}}} \, .
\end{equation}
This characterises the degeneracies between different parameters: $r=0$ means $p_i$ and $p_j$ are uncorrelated, while $r = \pm 1$ means they are completely (anti)correlated.

\subsection{Results}

Having applied the Fisher matrix $P(k,\mu)$ formalism described in the previous Section, we are now ready to present our results. In the following, we use \autoref{eq:Fish2} with $P^{\rm S}$ given by the TNS and EFTofLSS model at redshifts $z=0.5$ and $z=1$. As in \autoref{sec:sims-comparison}, we use $k_{\rm min}=0.006 \, h/{\rm Mpc}$ in all cases. Our fiducial model parameters and $k_{\rm max}$ are taken from \autoref{fittable}, and the survey parameters are the same as those of PICOLA simulations, namely survey (bin) volume $V_{\rm s}=4 \, {\rm Gpc}^3/h^3$ and number density of galaxies 
$n = 1 \times 10^{-3} \, h^3/{\rm Mpc}^3$. 

\subsubsection{TNS-based model forecasts}

\begin{figure}
\centering
  \includegraphics[scale=0.5]{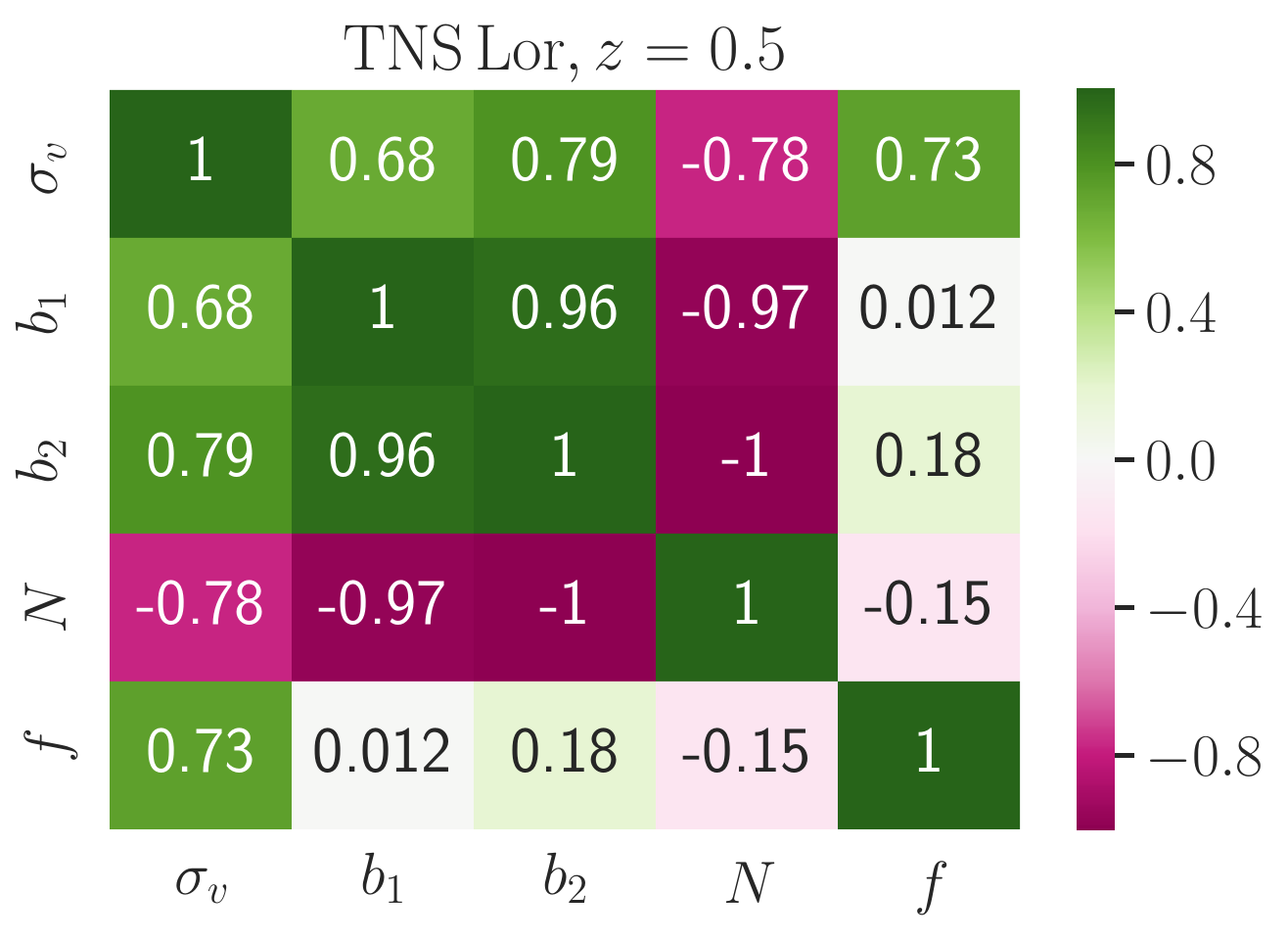}  
   \includegraphics[scale=0.5]{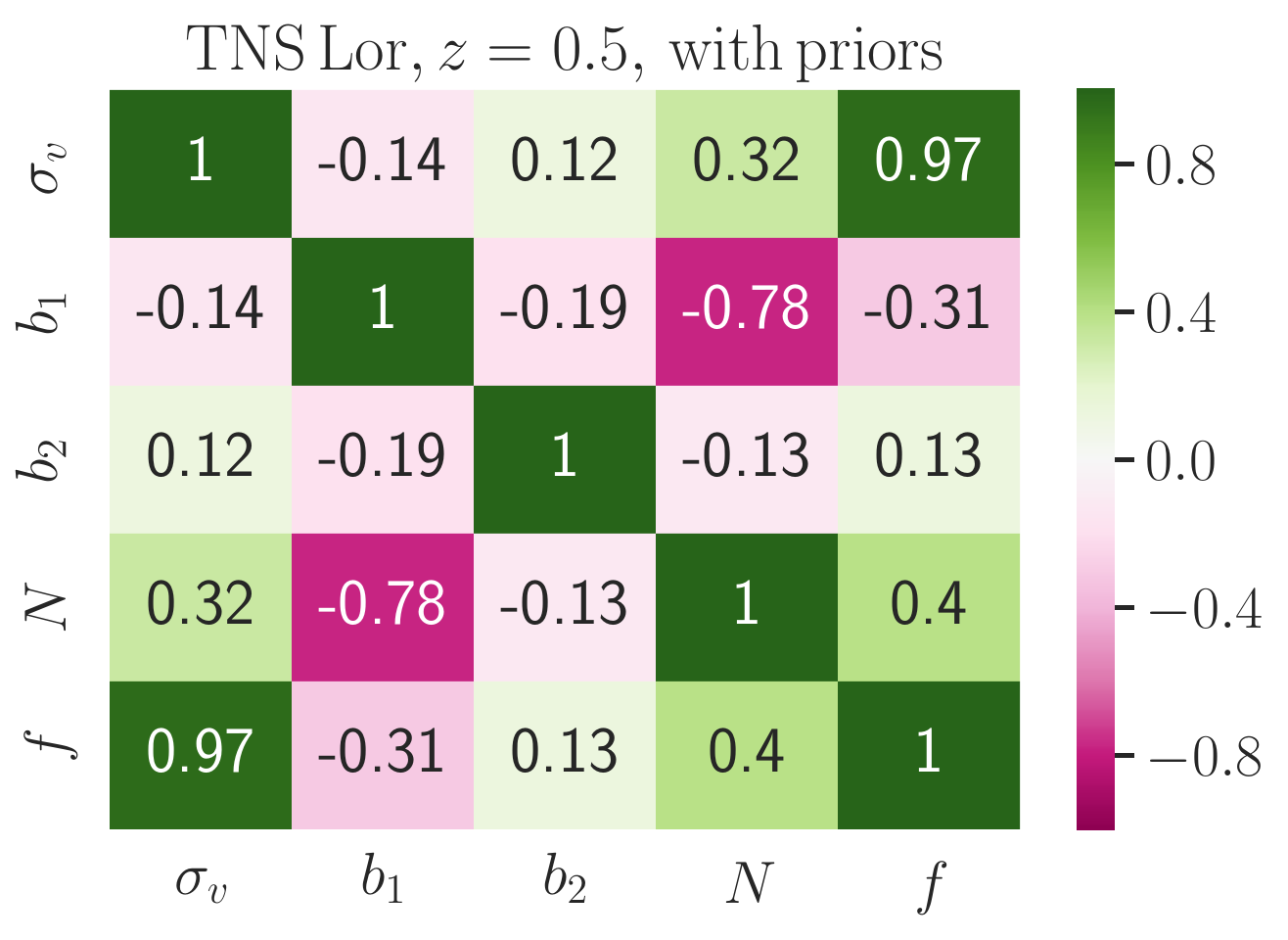}
  \caption[]{Correlation coefficient $r$ for the TNS model parameters of \autoref{redshiftps}) at $z=0.5$ with $k_{\rm max}=0.227 \, h$/Mpc. We show the results without any priors (top) and adding $10\%$ priors (bottom) on the nuisance parameters $\{\sigma_v,b_1,b_2,N\}$ as described in the main text.}
\label{fig:corr-mat-tns}
\end{figure}

\begin{figure}
\centering
 \includegraphics[scale=0.4]{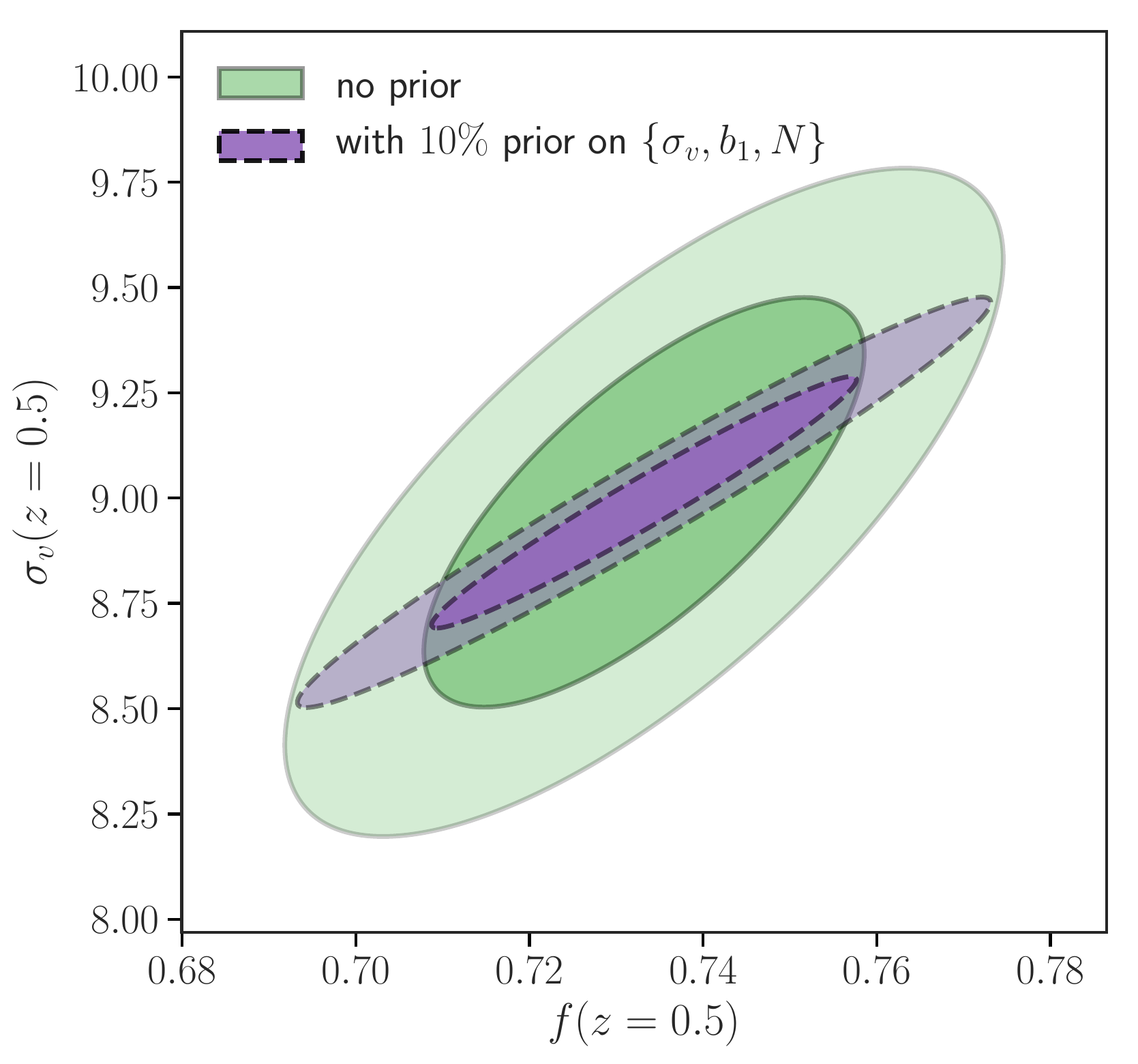}
   \includegraphics[scale=0.4]{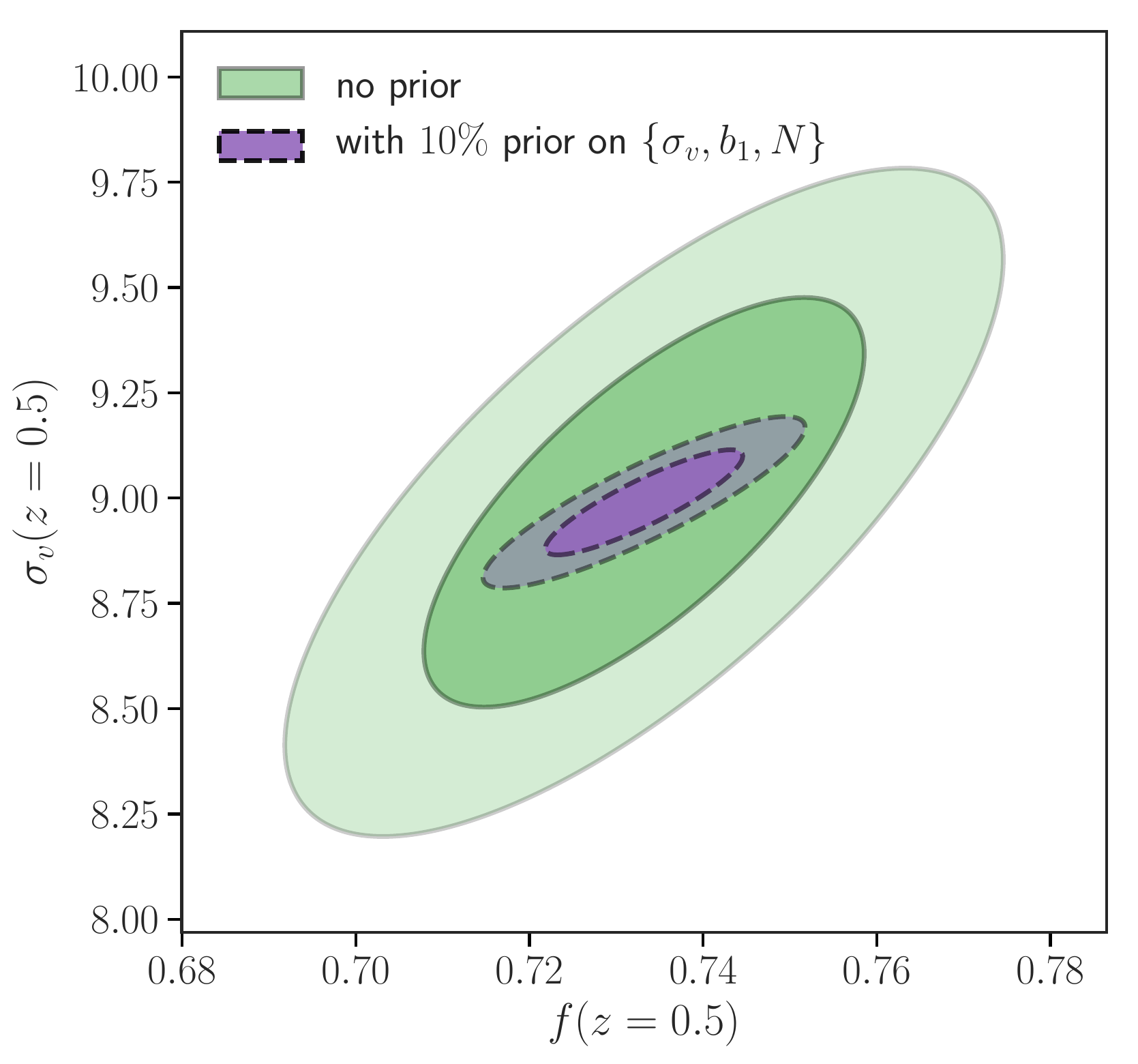}
  \caption[]{TNS model $1\sigma$ and $2\sigma$ confidence contours for $(f,\sigma_v)$, with and without the selected priors on $\{\sigma_v,b_1,N\}$ as described in the main text, at redshift $z=0.5$.}
\label{fig:contours-tns-zeq0p5}
\end{figure}

\begin{figure}
\centering
  \includegraphics[scale=0.5]{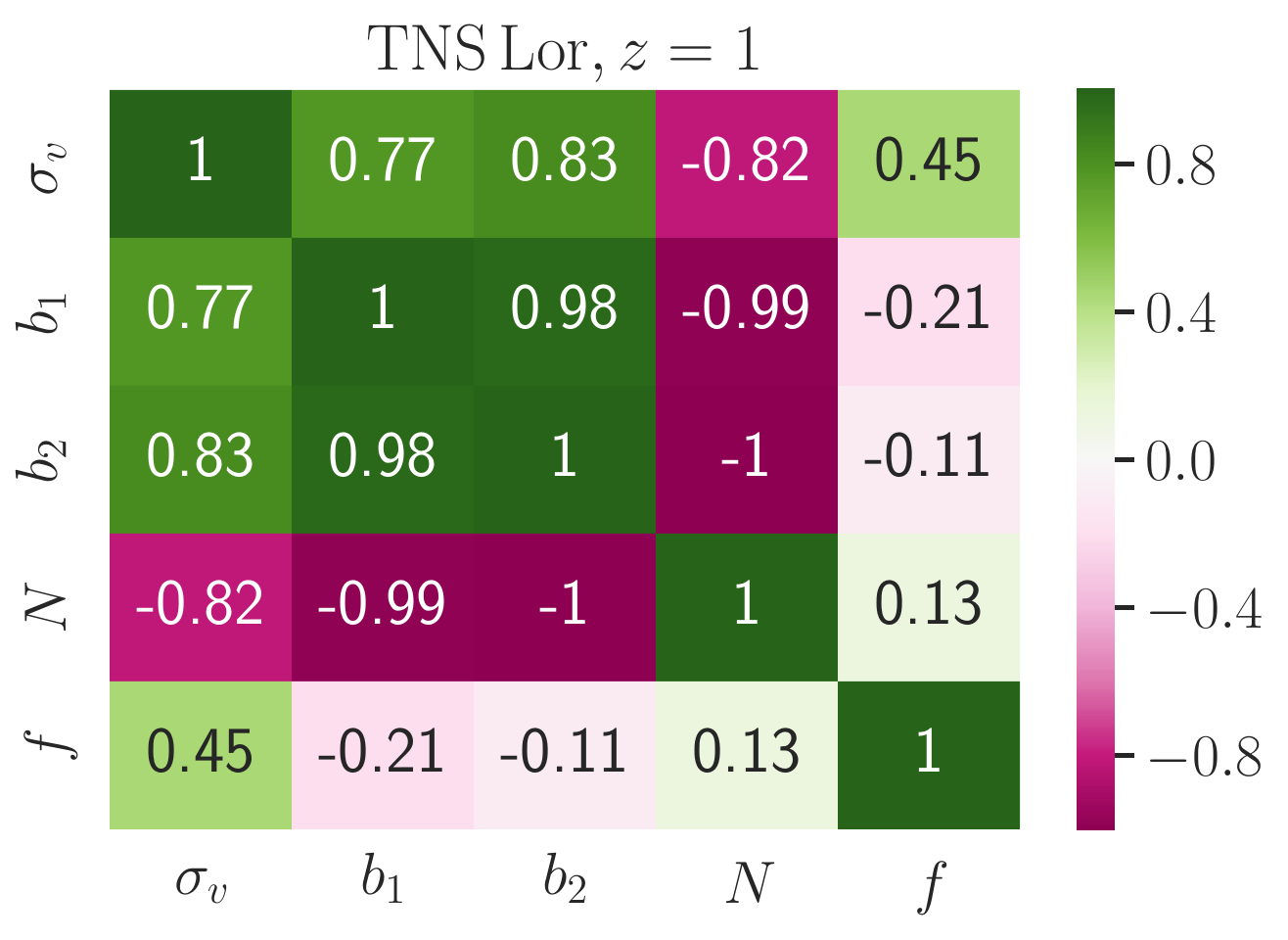}  
   \includegraphics[scale=0.5]{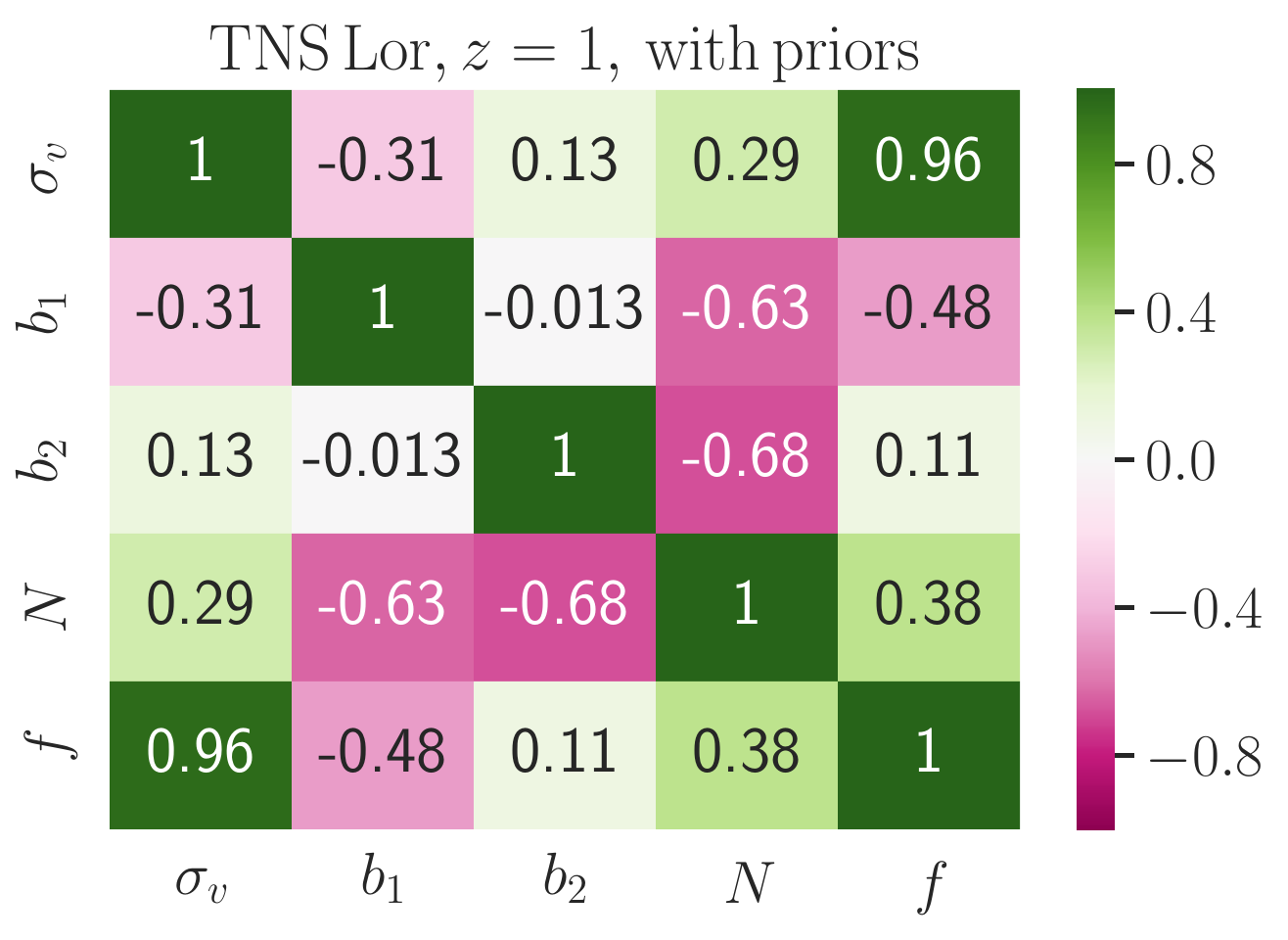}
  \caption[]{Correlation coefficient $r$ for the TNS model parameters of \autoref{redshiftps} at $z=1$ with $k_{\rm max}=0.276 \, h$/Mpc. We show the results without any priors (top) and with $10\%$ priors (bottom) on the nuisance parameters $\{\sigma_v,b_1,b_2,N\}$ as described in the main text.}
\label{fig:corr-mat-tns-zeq1}
\end{figure}

\begin{figure}
\centering
 \includegraphics[scale=0.45]{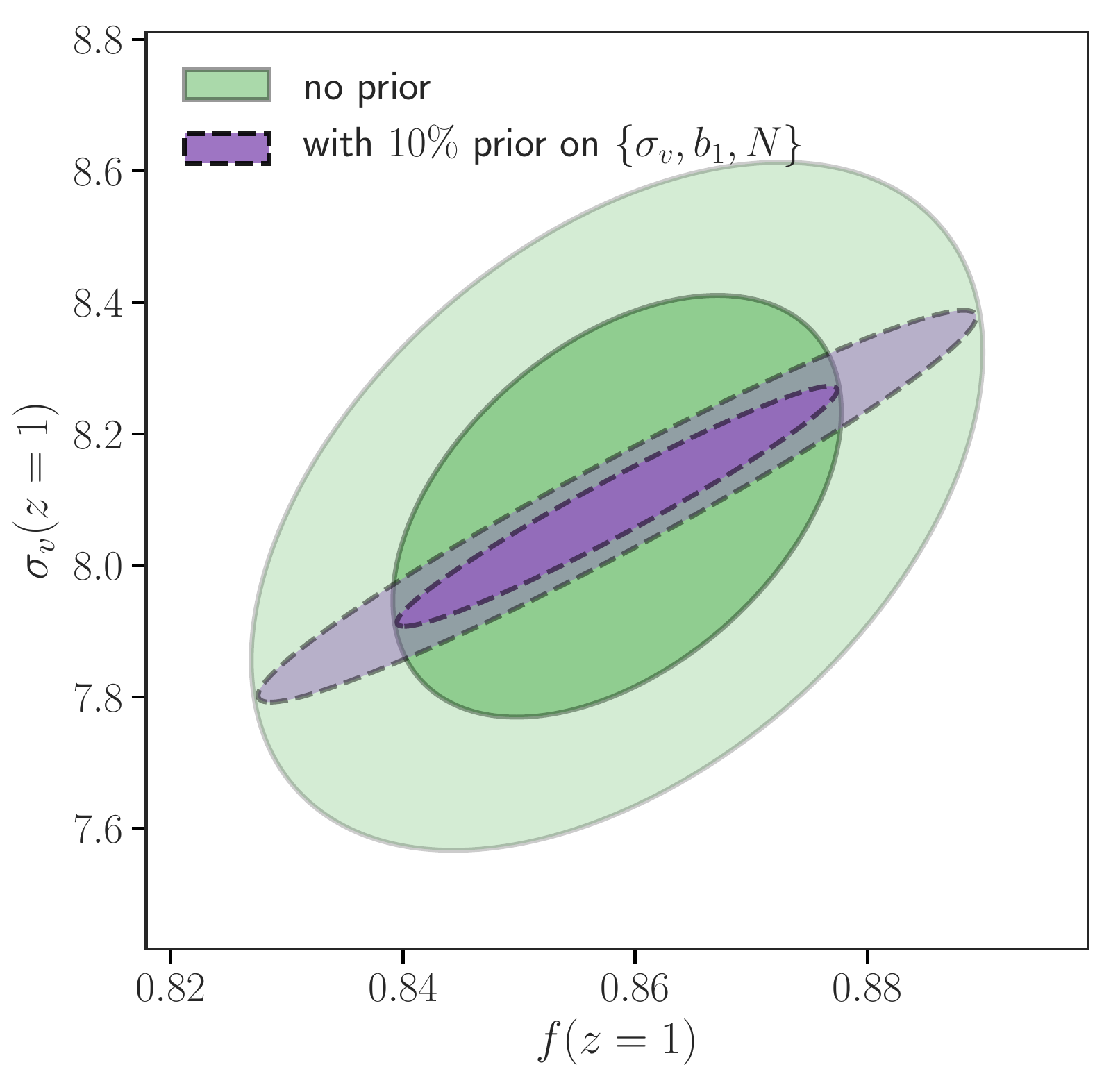}
   \includegraphics[scale=0.45]{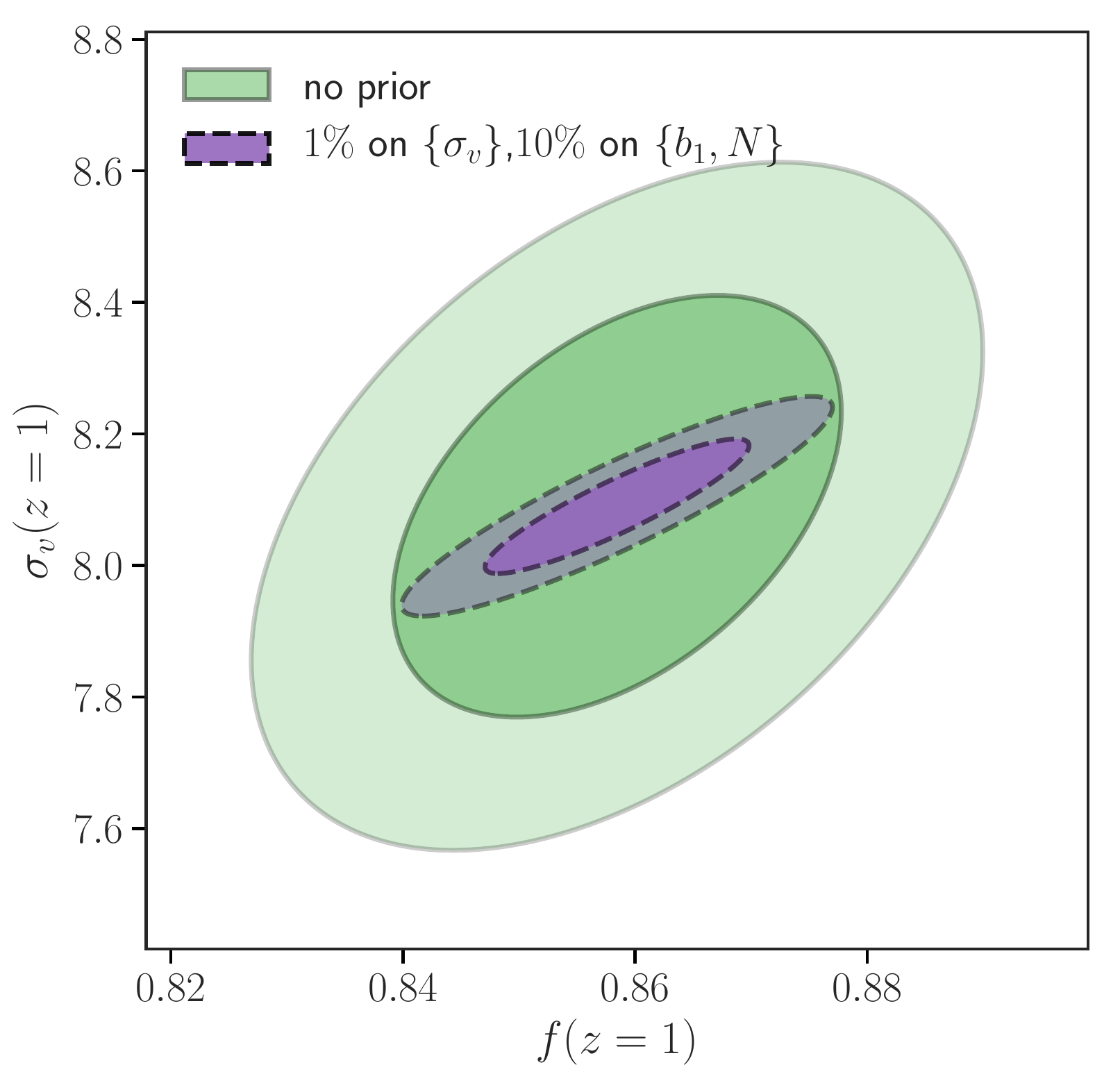}
  \caption[]{TNS model $1\sigma$ and $2\sigma$ confidence contours for $(f,\sigma_v)$,  with and without the selected priors on $\{\sigma_v,b_1,N\}$ as described in the main text, at redshift $z=1$.}
\label{fig:contours-tns-zeq1}
\end{figure}

We are first going to work with the TNS-based model in \autoref{redshiftps} with a Lorentzian $D_{\rm FoG}$; we will not consider the Gaussian FoG since it performs considerably worse, as discussed in \autoref{sec:sims-comparison}.
We are going to vary the parameters  $\{\sigma_v,b_1,b_2,N,f\}$ in two redshift bins of equal volume centred at $z=0.5$ and $z=1$. As we have already mentioned, the first four parameters, $\{\sigma_v,b_1,b_2,N\}$ are the model's nuisance parameters, and $f$ is the growth of structure. This is the cosmological parameter we are mainly interested in measuring with Stage IV surveys. We also want to investigate important questions regarding the use of these models for analysis when Stage IV data become available: for example, it is crucial to investigate the degeneracies between cosmological parameters of interest and additional nuisance parameters (needed to model the small scales), as well as the effects of priors. 
\\\\
Let us start with the results at $z=0.5$. Here, from \autoref{fittable} we have the fiducial values for all the parameters and $k_{\rm max} = 0.227 \, h/{\rm Mpc}$. We begin by letting all the parameters vary without imposing any priors. We perform the Fisher matrix analysis and show the resulting correlation coefficient matrix in \autoref{fig:corr-mat-tns} (top). As we can see, there are significant correlations between several of the model parameters $\{\sigma_v,b_1,b_2,N\}$, and between $\sigma_v$ and the cosmological parameter $f$.  
We find that the final $1\sigma$ percentage error on the structure growth rate $f$, marginalised over all other parameters,  is $\simeq 2.3 \% $. 
\\\\
The constraints can be improved if we put a prior on the model's nuisance parameters. Imposing a $10\%$ Gaussian prior across $\{\sigma_v,b_1,b_2,N\}$ results in some significant decorrelations, as demonstrated in \autoref{fig:corr-mat-tns} (bottom). 
The final $1\sigma$ percentage error on the structure growth $f$, marginalised over all other parameters, is reduced to $\simeq 1.9 \% $. 
That is, a $10\%$ prior on the nuisance parameters results in a $\sim 20\%$ improvement in the measurement of $f$ at $z=0.5$. In other words, as expected, if we let the nuisance parameters to be determined solely from the data at hand, jointly with the cosmological parameter without any priors, the constraint on $f$ is weakened due to the additional degeneracies.  
\\\\
However, imposing a $10\%$ prior across all the TNS nuisance parameters is not realistic. For example, it is very difficult to get an independent measurement of $b_2$ at this level. Importantly, a prior on the other three parameters $\{\sigma_v,b_1,N\}$ at the $10\%$ level is much more realistic: $b_1$ can be constrained using additional information from the bispectrum (see \citet{Yankelevich:2018uaz} for Euclid-like forecasts), $N$ can be measured, and $\sigma_v$'s degeneracy with $f$ can be broken by additional modelling, as well as priors motivated by simulations and/or halo model predictions \citep[e.g.][]{Zheng:2016zxc,Zheng:2016xvo}. We therefore proceed to present constraints imposing a $10\%$ prior across $\{\sigma_v,b_1,N\}$. In \autoref{fig:contours-tns-zeq0p5} (top), we show the $1\sigma$ and $2\sigma$ confidence contours for the $(f,\sigma_v)$ parameters at $z=0.5$, with and without this prior. The percentage error on $f$ is $2.2\%$, and it becomes evident that the constraint on $f$ will significantly improve with a stronger prior on $\sigma_v$. As illustration, imposing a $1\%$ prior on $\sigma_v$ we indeed find that the percentage error on $f$ is reduced to $1\%$, and the confidence contours are shown in \autoref{fig:contours-tns-zeq0p5} (bottom).
\newline
\newline
\noindent
We will now present the results at $z=1$. Here, from \autoref{fittable} we have the fiducial values for all the parameters and $k_{\rm max} = 0.276 \, h/{\rm Mpc}$. We follow the same procedure as before, i.e. first letting all the parameters vary freely, and then imposing a $10\%$ Gaussian prior across $\{\sigma_v,b_1,b_2,N\}$. We show the resulting correlation coefficient matrices for $z=1$ in \autoref{fig:corr-mat-tns-zeq1}. The final $1\sigma$ percentage error on the structure growth $f$, marginalised over all other parameters,  is $\simeq 1.5 \% $. This is smaller than the fractional error for $f$ we obtained at $z=0.5$, mainly because of the significantly higher $k_{\rm max}$ at this redshift. Including the $10\%$ priors, the constraint on $f$ is reduced to $\simeq 1.4 \% $, a marginal improvement. 
\\\\
Following the same reasoning as before, we present results imposing a $10\%$ prior on  $\{\sigma_v,b_1,N\}$, and then making the prior on $\{\sigma_v\}$ much stronger, $1\%$. The former results in a $\simeq 1.5\%$ error on $f$, while the latter reduces the error to $\simeq 0.9\%$. The confidence contours for $(f,\sigma_v)$ at $z=1$ with and without the imposed priors are shown in  \autoref{fig:contours-tns-zeq1}. 


\subsubsection{EFTofLSS-based model forecasts}

\begin{figure}
\centering
  \includegraphics[scale=0.45]{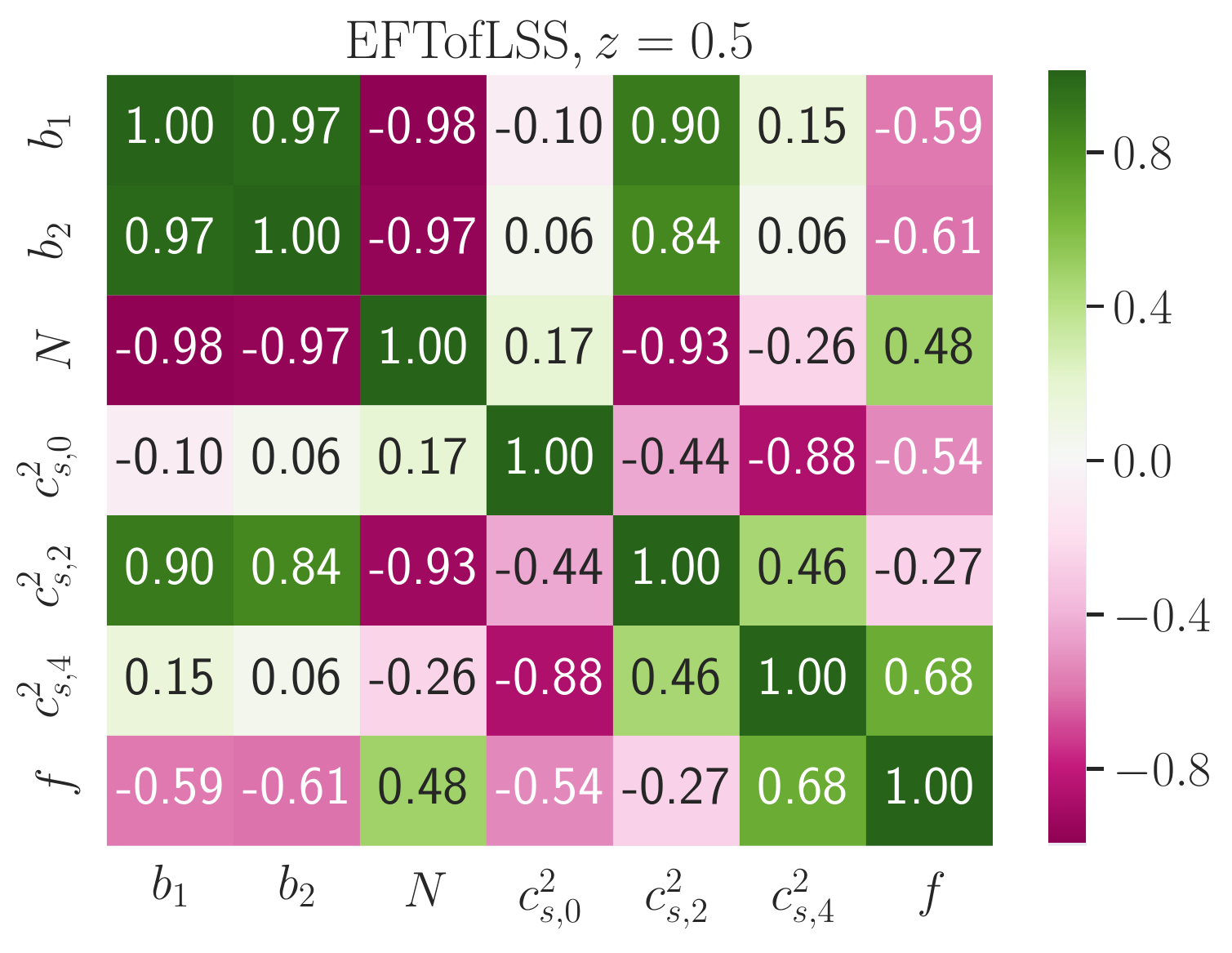}  
   \includegraphics[scale=0.45]{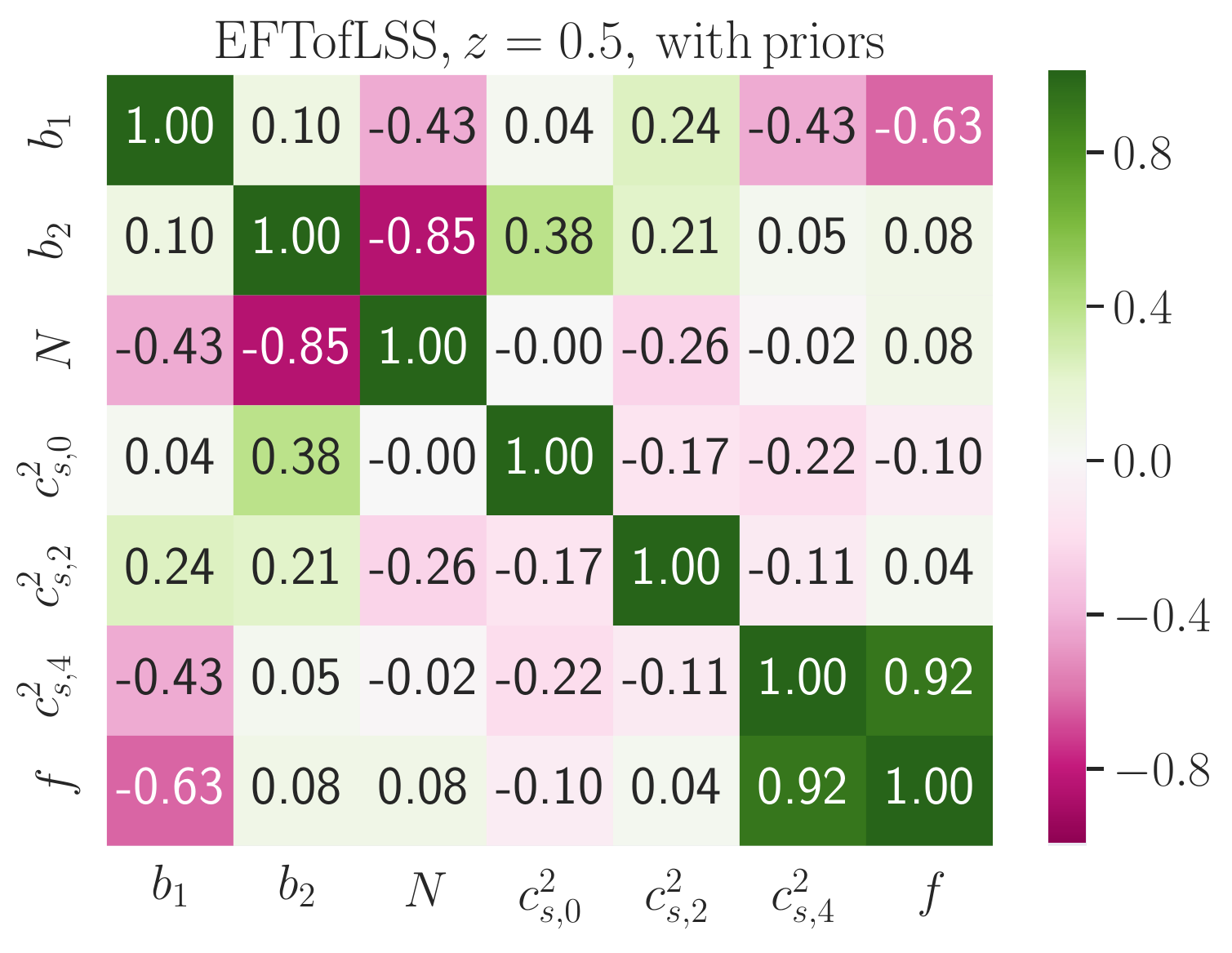}
  \caption[]{Correlation coefficient $r$ for the EFTofLSS model parameters of \autoref{redshiftps2} at $z=0.5$ with $k_{\rm max}=0.245 \, h$/Mpc. We show the results without any priors (top) and with $10\%$ priors (bottom) on the nuisance parameters $\{b_1,b_2,N,c^2_{s,0},c^2_{s,2},c^2_{s,4}\}$ as described in the main text.}
\label{fig:corr-mat-eft}
\end{figure}

We now move on to the EFTofLSS-based model, \autoref{redshiftps2}.
The set of parameters we are going to vary is $\{b_1,b_2,N,c^2_{s,0},c^2_{s,2},c^2_{s,4},f\}$, for the two redshift bins centred at $z=0.5$ and $z=1$.
\\\\
We start with the results at $z=0.5$. Here, from \autoref{fittable} we have the fiducial values for all the parameters and $k_{\rm max} = 0.245 \, h/{\rm Mpc}$. Following our TNS-based model analysis presented before, we begin by letting all the parameters vary without imposing any priors. 
We perform the Fisher matrix analysis and show the resulting correlation coefficient matrix in \autoref{fig:corr-mat-eft} (top). 
Again, there are significant correlations between several parameters. 
We find that the final $1\sigma$ percentage error on the structure growth $f$, marginalised over all other parameters,  is $\simeq 3.3 \% $;
which is worse than the TNS-based model at this redshift (that gave $2.3\%$), despite the higher $k_{\rm max}$ at this redshift. 
Imposing a $10\%$ Gaussian prior across $\{b_1,b_2,N,c^2_{s,0},c^2_{s,2},c^2_{s,4}\}$ results in some significant decorrelations, as demonstrated in \autoref{fig:corr-mat-eft} (bottom). 
The final $1\sigma$ percentage error on the structure growth $f$, marginalised over all other parameters, is reduced to $\simeq 1.8 \% $. This is a major improvement, but imposing such priors on all the nuisance EFTofLSS parameters is not realistic. 
\\\\
Imposing a $10\%$ Gaussian prior on the parameters $\{b_1,N\}$ is more conservative, and the error on $f$ using this prior is $\simeq 2.9 \% $. This result demonstrates that the degeneracies brought by the $\{c^2_{s,0},c^2_{s,2},c^2_{s,4}\}$ EFTofLSS parameters are significant. Note that priors on these parameters at a given redshift can be obtained if we can predict their time dependence from theory \citep{Foreman:2015uva}, in combination with a measurement at some other redshift. We show the $1\sigma$ and $2\sigma$ confidence contours for the parameters $(f,N)$ at $z=0.5$ in \autoref{fig:contours-eft} (top).
\noindent
\newline
\newline
We will now present the results at $z=1$. Here, from \autoref{fittable} we have the fiducial values for all the parameters and $k_{\rm max} = 0.276 \, h/{\rm Mpc}$. We follow the same procedure as before.\footnote{ In the EFTofLSS case without any priors we find that the $c^2_{s,0}$ parameter can take negative values. A way to mitigate this is to impose a prior on this parameter. Note that due to the nature of the Fisher matrix formalism, this prior cannot be flat; it has to be Gaussian and hence we cannot completely avoid the occurrence of negative values, but we can make them far less likely. This also means that we artificially make the possibility of large positive values less likely. Since the fiducial $c^2_{s,0}$ value from \autoref{fittable} is practically zero at $z=1$, and this Fisher analysis is mainly exploratory, we choose not to impose a prior and we let the parameter free to vary (we do the same at $z=0.5$ for consistency). We will return to this issue when we perform the Fisher matrix and MCMC comparison in \autoref{sec:FishMCMCcomp}.} We show the resulting correlation coefficient matrices for $z=1$ in \autoref{fig:corr-mat-eft-zeq1}. The final $1\sigma$ percentage error on the structure growth $f$, marginalised over all other parameters,  is $\simeq 3.1\% $. Including the $10\%$ priors across all nuisance parameters, the constraint on $f$ is reduced to $\simeq 1.7\% $. 
Imposing the moderate $10\%$ prior on the $\{b_1,N\}$ parameters only, we find that the error on $f$ is $\simeq 2.8\%$ -- this is again worse than the results of the TNS-based model with the imposed conservative priors at $z=1$. We show the $1\sigma$ and $2\sigma$ confidence contours for the parameters $(f,N)$ at $z=1$ in \autoref{fig:contours-eft} (bottom). 
\newline
\newline
 It is important to note that the $k_{\rm max}$ found here is very high compared to previous studies. For example in the BOSS analysis, the $k_{\rm max}= 0.15 \, h/{\rm Mpc}$ at $z=0.61$. Furthermore, the dark-matter-only TNS model is only able to fit up to around $k\sim 0.2 \, h/{\rm Mpc}$ at $z=1$ (e.g \cite{Taruya:2010mx,Bose:2017myh}). This suggests that the bias model adopted here is accounting for a break down of the RSD model. Such considerations prompt us to investigate the constraints and parameter degeneracies at a more conservative $k_{\rm max}$ next. 
\\
\\

\subsubsection{Conservative forecasts}

\begin{figure}
\centering
  \includegraphics[scale=0.45]{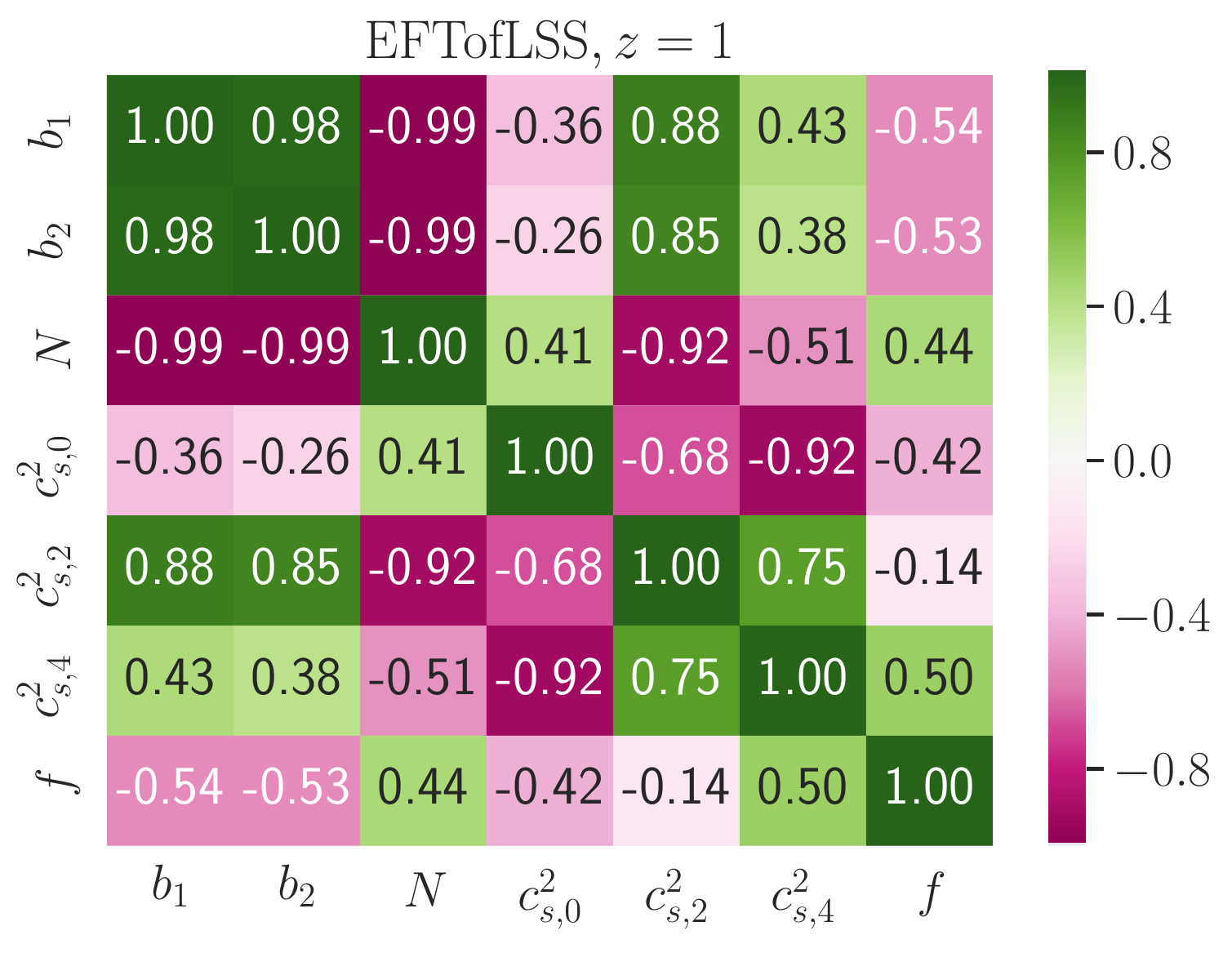}  
   \includegraphics[scale=0.45]{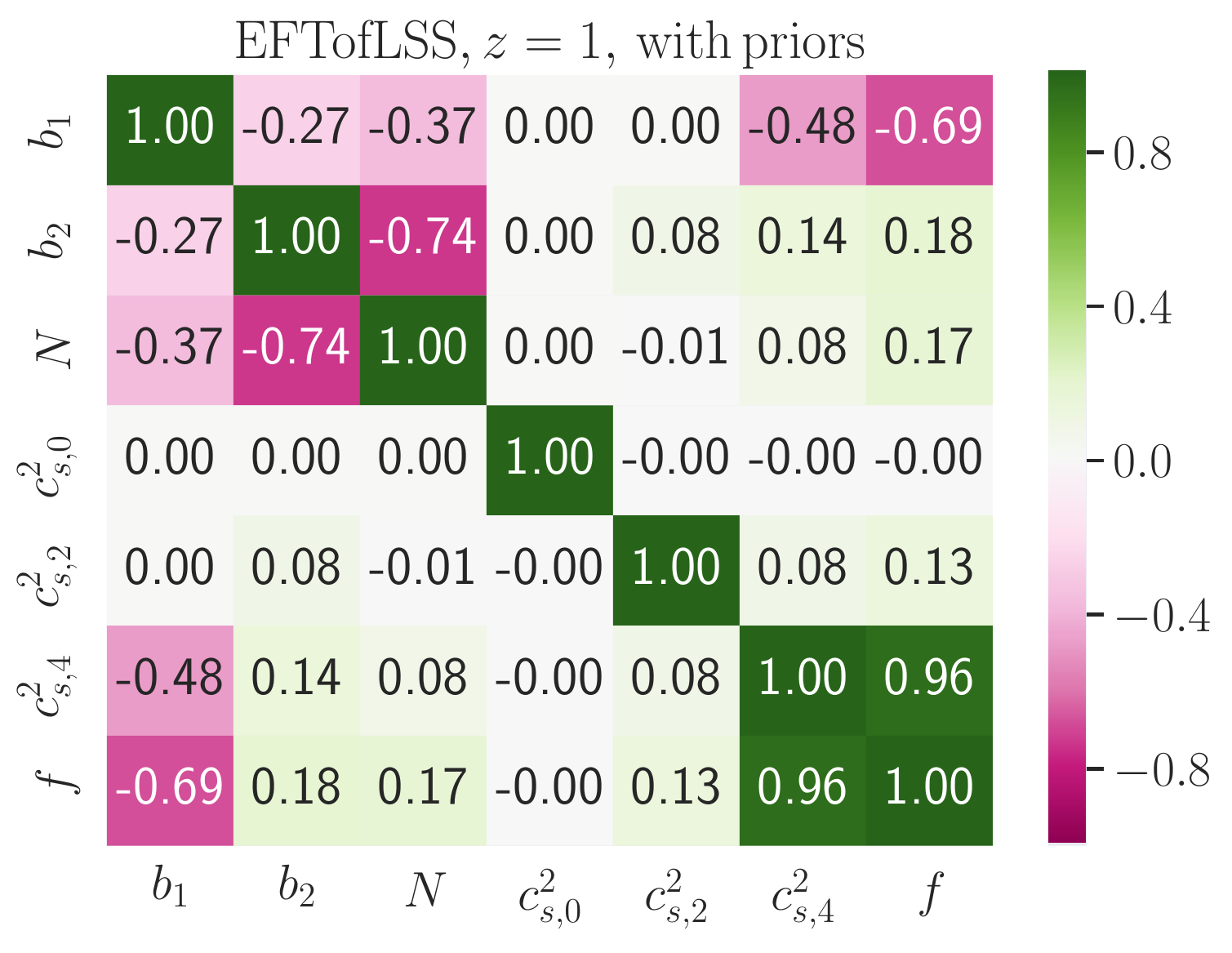}
  \caption[]{Correlation coefficient $r$ for the EFTofLSS-based model parameters of \autoref{redshiftps2} at $z=1$ with $k_{\rm max}=0.276 \, h$/Mpc. We show the results without any priors (top) and with $10\%$ priors (bottom) on the nuisance parameters $\{b_1,b_2,N,c^2_{s,0},c^2_{s,2},c^2_{s,4}\}$ as described in the main text.}
\label{fig:corr-mat-eft-zeq1}
\end{figure}

\begin{figure}
\centering
 \includegraphics[scale=0.4]{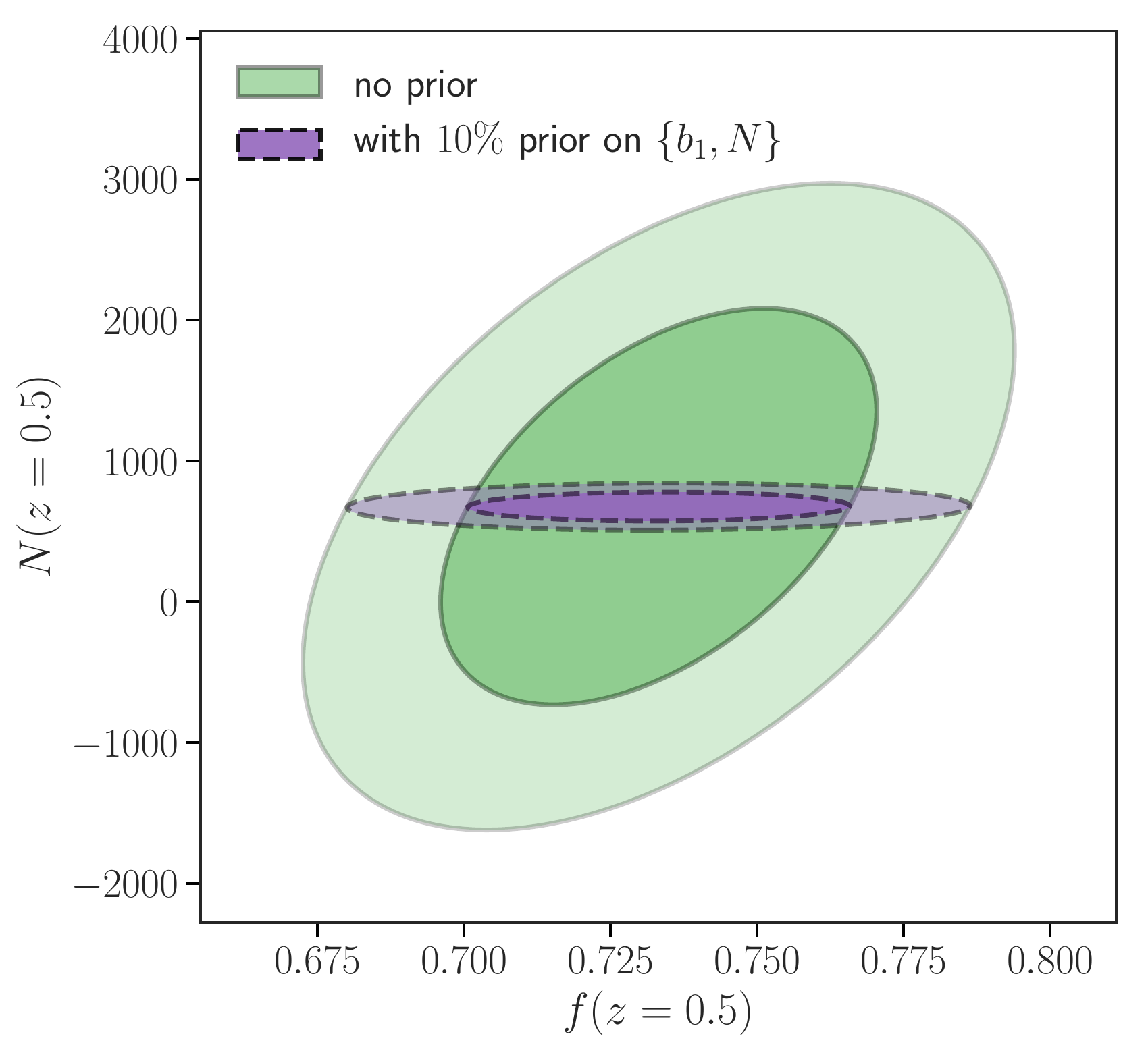}
   \includegraphics[scale=0.4]{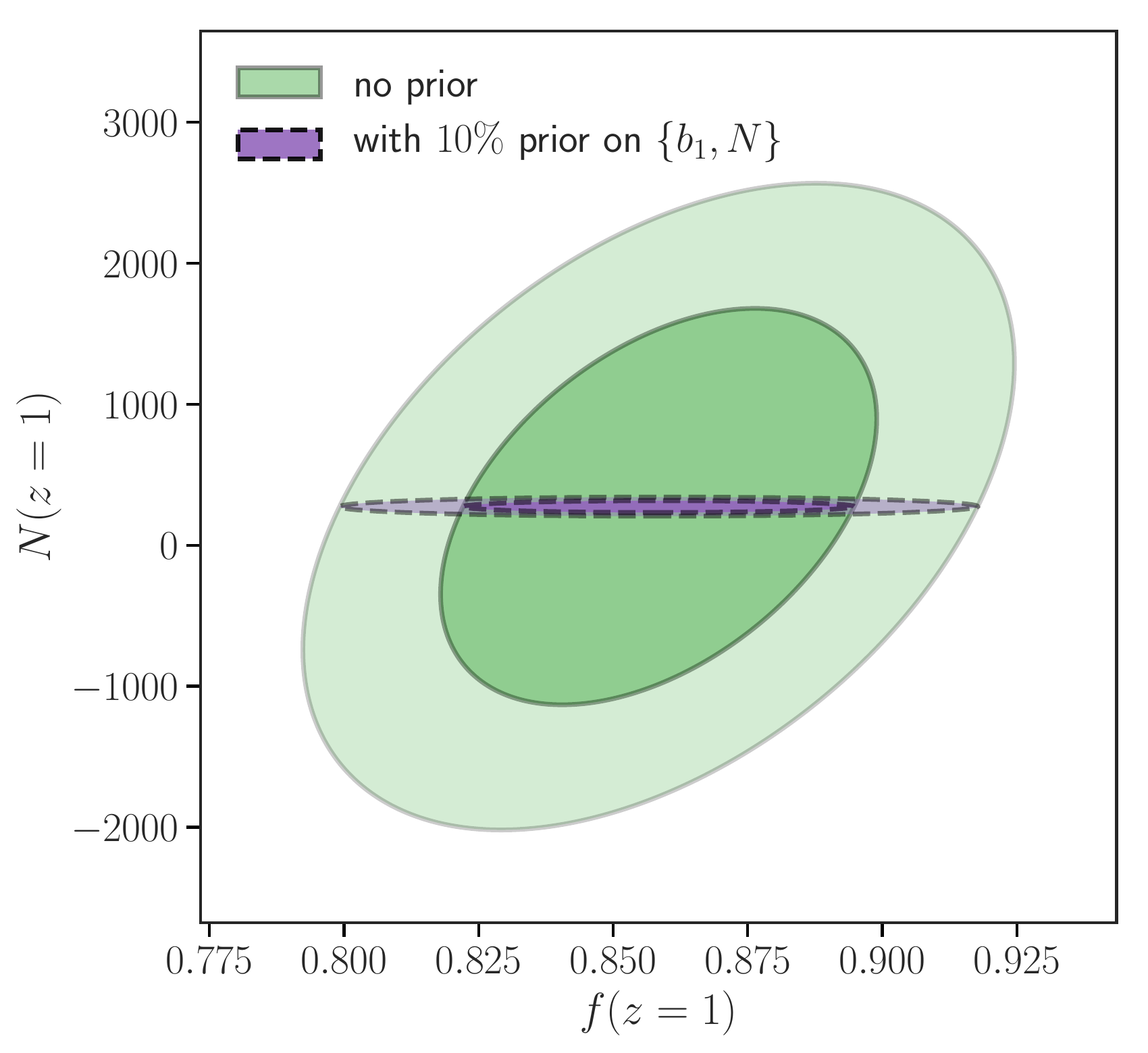}
  \caption[]{EFTofLSS-based model $1\sigma$ and $2\sigma$ confidence contours for $(f,N)$, with and without the selected priors on $\{b_1,N\}$ as described in the main text, at redshifts $z=0.5$ (top) and $z=1$ (bottom).}
\label{fig:contours-eft}
\end{figure}

\begin{figure}
\centering
  \includegraphics[scale=0.5]{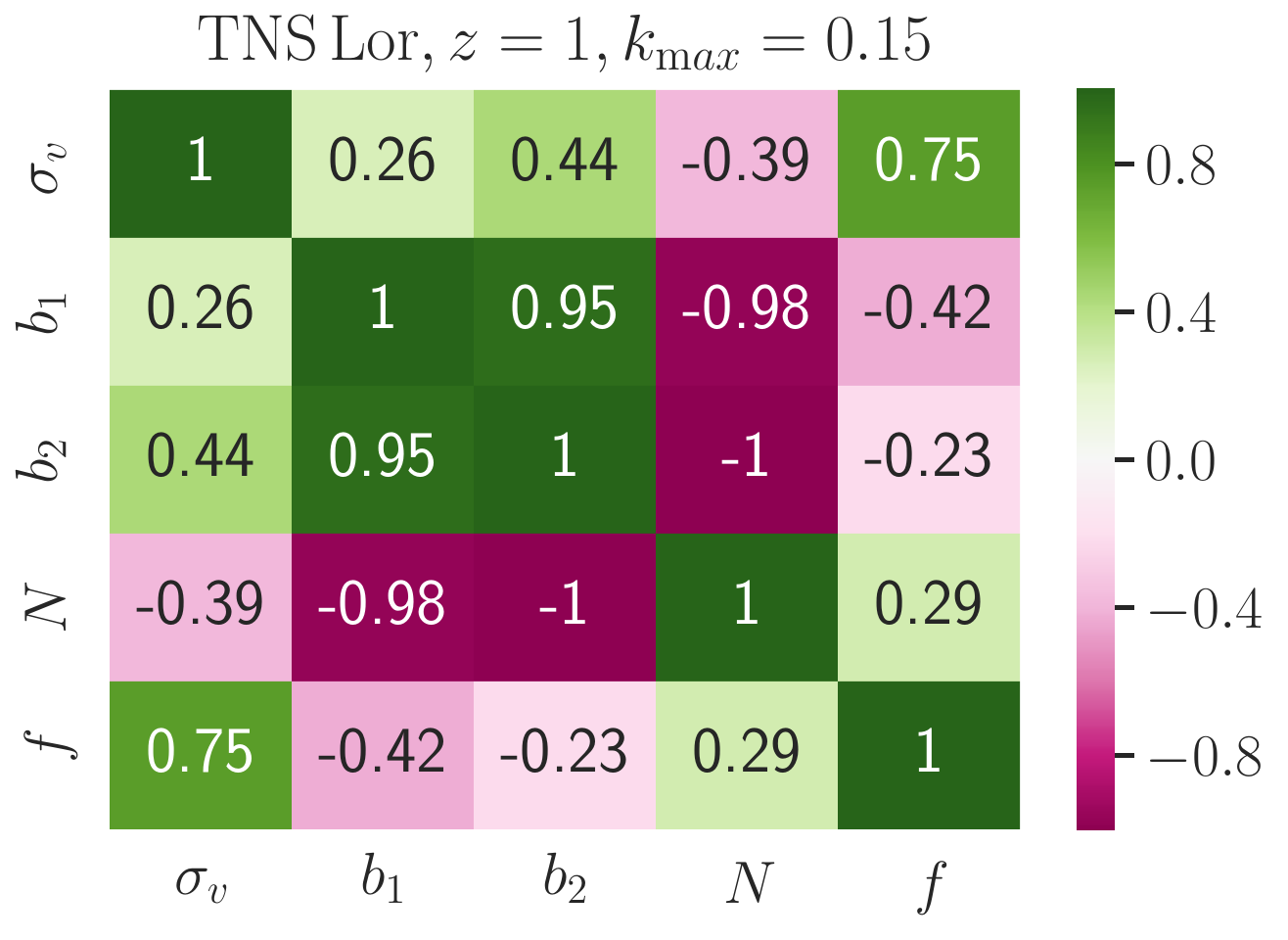}  
  \includegraphics[scale=0.45]{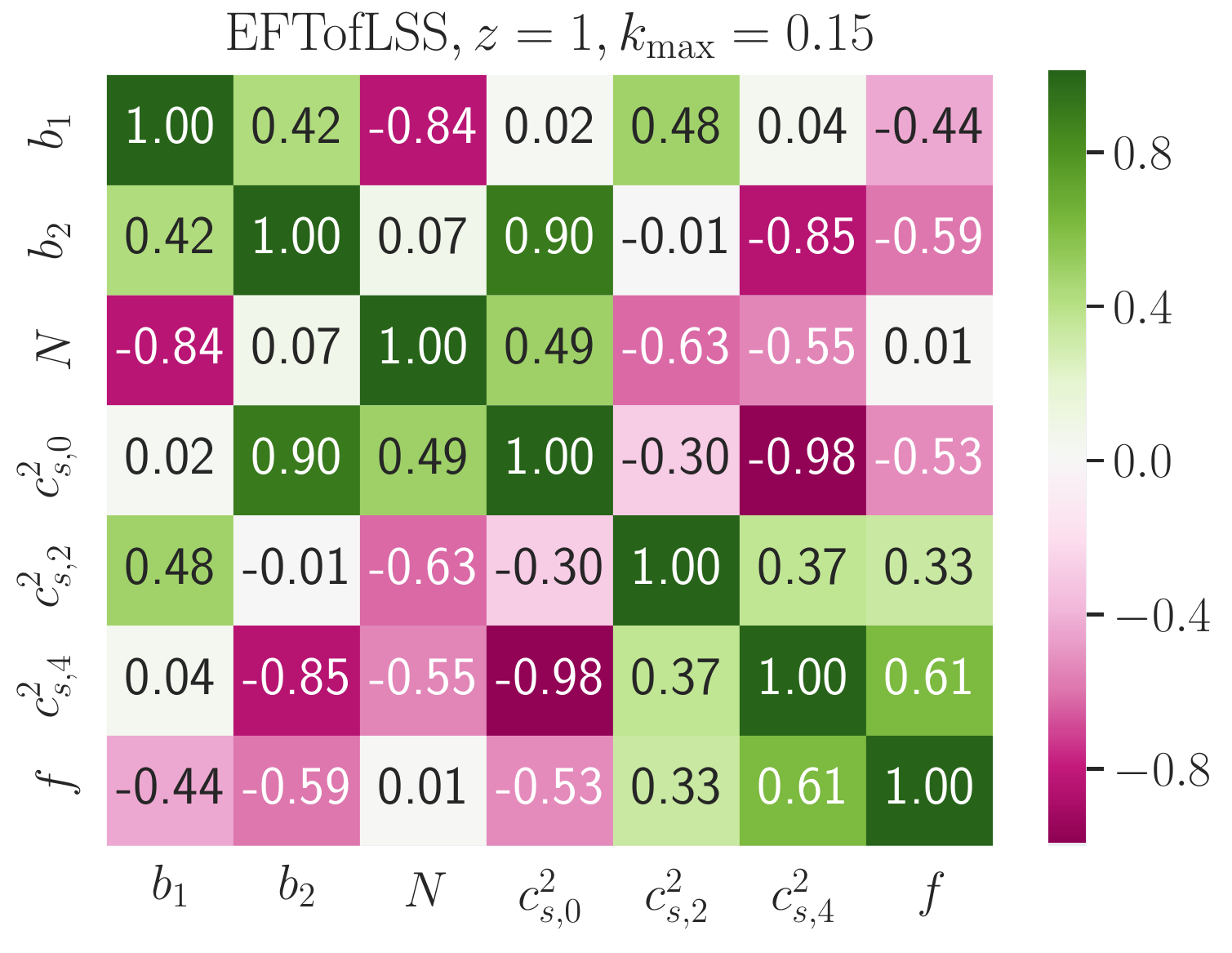}
  \caption[]{Correlation coefficient $r$ for the TNS (top) and EFTofLSS (bottom) parameters at $z=1$ with a conservative $k_{\rm max}=0.15 \, h$/Mpc.}
\label{fig:corr-mat-tns-zeq1-cons-kmax}
\end{figure}

An interesting result of the analysis summarised in \autoref{fittable} is that both the TNS and EFTofLSS models have the same $k_{\rm max}=0.276 \, h$/Mpc at $z=1$. In this subsection we focus on this redshift to explore how the parameter degeneracies and constraints change if we assume a much more conservative $k_{\rm max}=0.15 \, h$/Mpc, the same for both models for the case of no priors. Starting with the TNS model, we show the correlation coefficient $r$ in the top panel of \autoref{fig:corr-mat-tns-zeq1-cons-kmax}. It is interesting to see how the various degeneracies change compared to \autoref{fig:corr-mat-tns-zeq1} (top panel). For example the $(f,\sigma_v)$ correlation has increased with respect to $k_{\rm max}=0.276 \, h$/Mpc, and the corresponding $f$ constraint has also increased from $2.3\%$ to $4.6\%$; this is expected due to the much smaller range of scales used. For EFTofLSS we show the correlation coefficient $r$ in the bottom panel of \autoref{fig:corr-mat-tns-zeq1-cons-kmax}. Significant decorrelations occur compared to \autoref{fig:corr-mat-eft-zeq1} (top panel). The $f$ constraint increases, from $3.1\%$ to $5.3\%$, as the range of scales is much smaller, but the comparison between the TNS and EFTofLSS $f$ constraints with this conservative $k_{\rm max}=0.15 \, h$/Mpc is more equalised.
\newline
\newline
  These results, summarised in \autoref{Fisherconstraintstab}, suggest that TNS is a better model prescription to use for future surveys, at both  $z\simeq 0.5$ and  $z\simeq 1$. However, as shown in \citetalias{Markovic:2019sva}, one has to use the multipoles analysis to get reliable forecasts, and we proceed to do this in the next Sections. 
First, we will move on to present an MCMC analysis using the monopole, quadrupole, and hexadecapole spectra. 
 An MCMC analysis is generally expected to be more reliable than Fisher matrix forecasts, as it can probe non-Gaussian posteriors and does not suffer from numerical instabilities that can sometimes be encountered in Fisher analyses \citep[e.g][]{Sprenger:2018tdb}. It also closely resembles a real data analysis procedure, and allows us to study biases on the estimation of the cosmological parameter of interest, $f$ \citepalias[e.g.][]{Markovic:2019sva}.
\begin{table*}
\centering
\caption{$1\sigma$ marginalised percent errors on $f$ from the Fisher analyses at $z=0.5$ and $z=1$. We use the full anisotropic power spectrum $P(k,\mu)$. The results correspond to the $k_{\rm max}$ values given in \autoref{fittable} for $z=0.5$ and $z=1$. We show results with and without selected moderate priors on $\{\sigma_v,b_1,N\}$ (TNS) and $\{b_1,N\}$ (EFTofLSS), as described in the main text.  We also show results with a more conservative $k_{\mathrm{max}}$, as described in the main text.}
\begin{tabular}{| c | l || c | c |}
\hline  
 \multicolumn{2}{|c||}{} & TNS-based model & EFTofLSS-based model \\
 \hline\hline
 \multirow{2}{*}[-1pt]{$z=0.5$} & $P(k,\mu)$ & $2.3\%$ & $3.3\%$ \\ \cline{2-4}
 & \rule{0pt}{3ex}$P(k,\mu)$ + $10\%$ prior& $2.2\%$  & $2.9\%$  \\[-2pt] 
 \hline\hline
\multirow{3}{*}[-2pt]{$z=1.0$} & $P(k,\mu)$ & $1.5\%$ & $3.1\%$  \\ \cline{2-4} 
 & \rule{0pt}{3ex}$P(k,\mu)$ + $10\%$ prior& $1.4\%$ & $2.8\%$   \\ \cline{2-4} 
 & \rule{0pt}{3ex} $P(k,\mu)|_{k_{\mathrm{max}}=0.15}$ &  $4.6\%$ &  $5.3\%$ \\[-2pt] \hline
\end{tabular}
\label{Fisherconstraintstab}
\end{table*}

\section{MCMC analysis} \label{sec:mcmc}
\begin{figure*}
\centering
  \includegraphics[height=0.59\textwidth,width=0.7\textwidth]{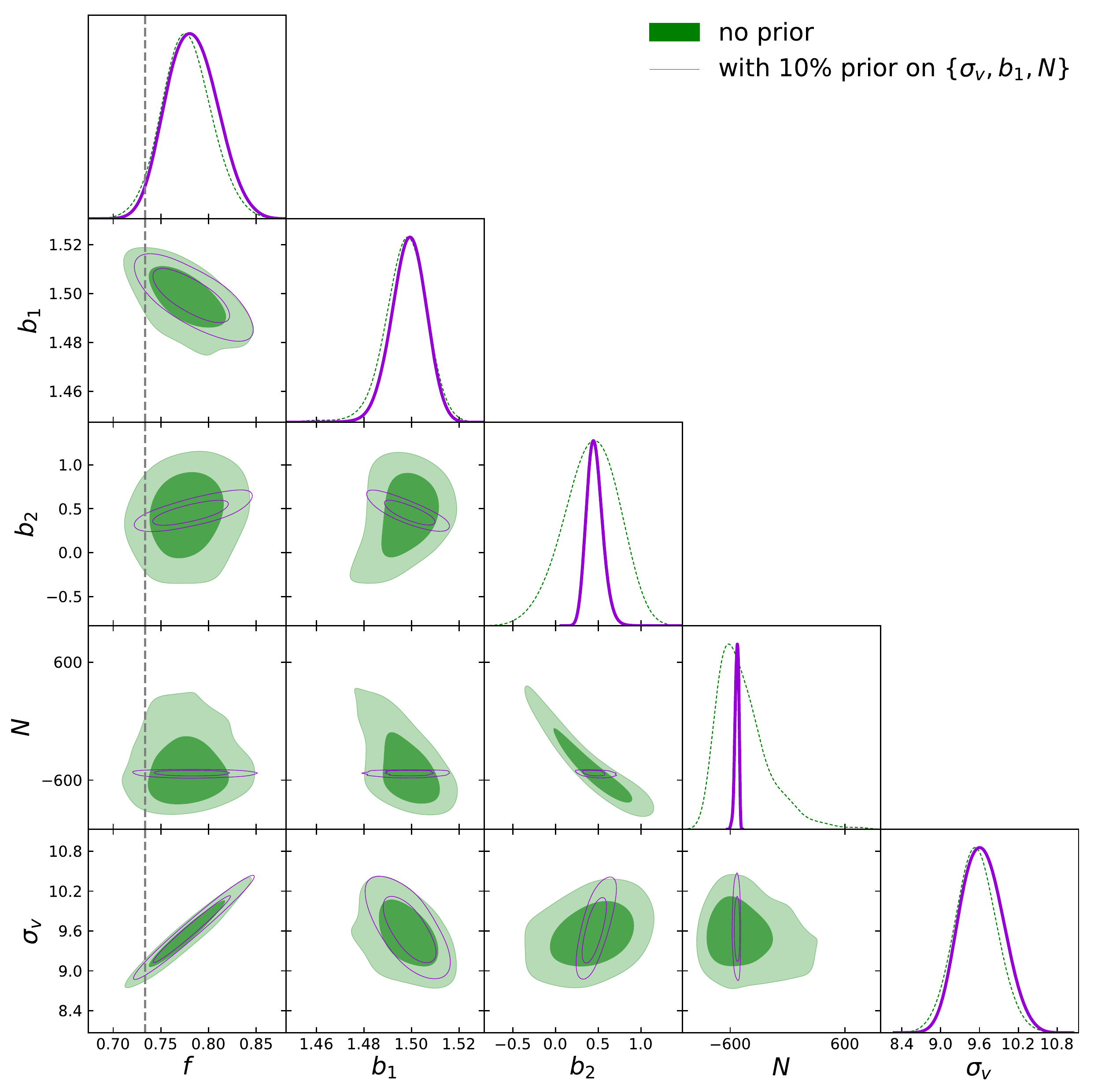}  
  \caption[]{MCMC results for the TNS model of \autoref{redshiftps} at $z=0.5$ with $k_{\rm max}~=~0.227~\,~h$/Mpc for $P_0$ and $P_2$ and $k_{\rm max,4}~=~0.129~\,~h$/Mpc for $P_4$. The dashed line indicates the fiducial value of $f=0.733$ in the PICOLA simulations. We also show the results with a $10\%$ prior around the best fit values of $\sigma_v,b_1$ and $N$. The best fit value of $f$ with $1 \sigma$ errors without (with) priors is $f=0.777\pm^{0.025}_{0.025}$ ($f=0.782\pm^{0.026}_{0.028}$).}
\label{mcmc1}
\end{figure*}

\begin{figure*}
\centering
  \includegraphics[height=0.8\textwidth,width=0.9\textwidth]{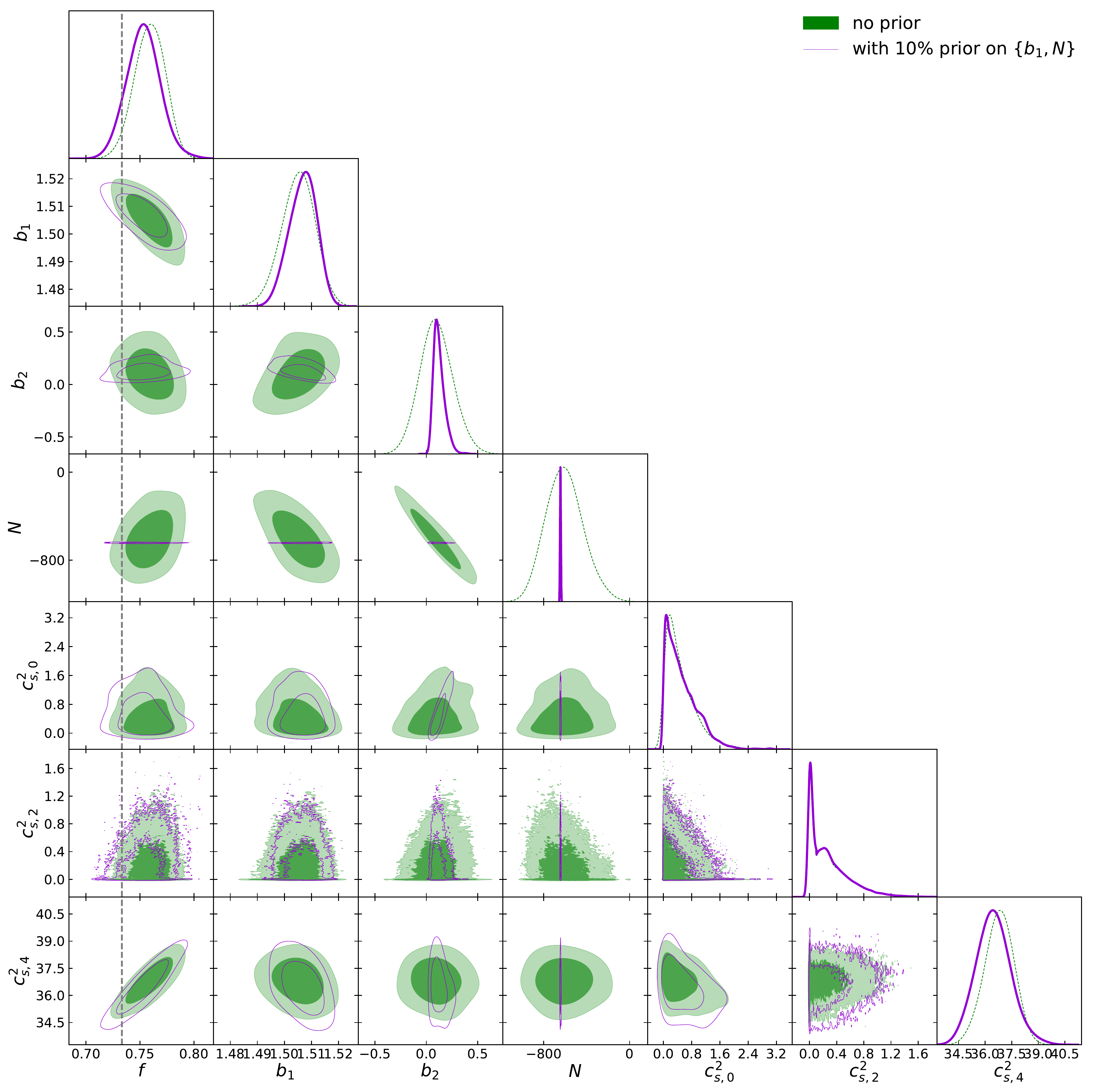}  
  \caption[]{MCMC results for the EFTofLSS model of \autoref{redshiftps2} at $z=0.5$ with $k_{\rm max}~=~k_{\rm max,4}~=~0.245~\,~h$/Mpc. The dashed line indicates the fiducial value of $f=0.733$ in the PICOLA simulations. We also show the results with a $10\%$ prior around the best fit values of $b_1$ and $N$. The best fit value of $f$ with $1 \sigma$ errors without (with) priors is $f=0.759\pm^{0.015}_{0.014}$ ($f=0.753\pm^{0.015}_{0.016}$).}
\label{mcmc2}
\end{figure*}

\begin{figure*}
\centering
  \includegraphics[height=0.59\textwidth,width=0.7\textwidth]{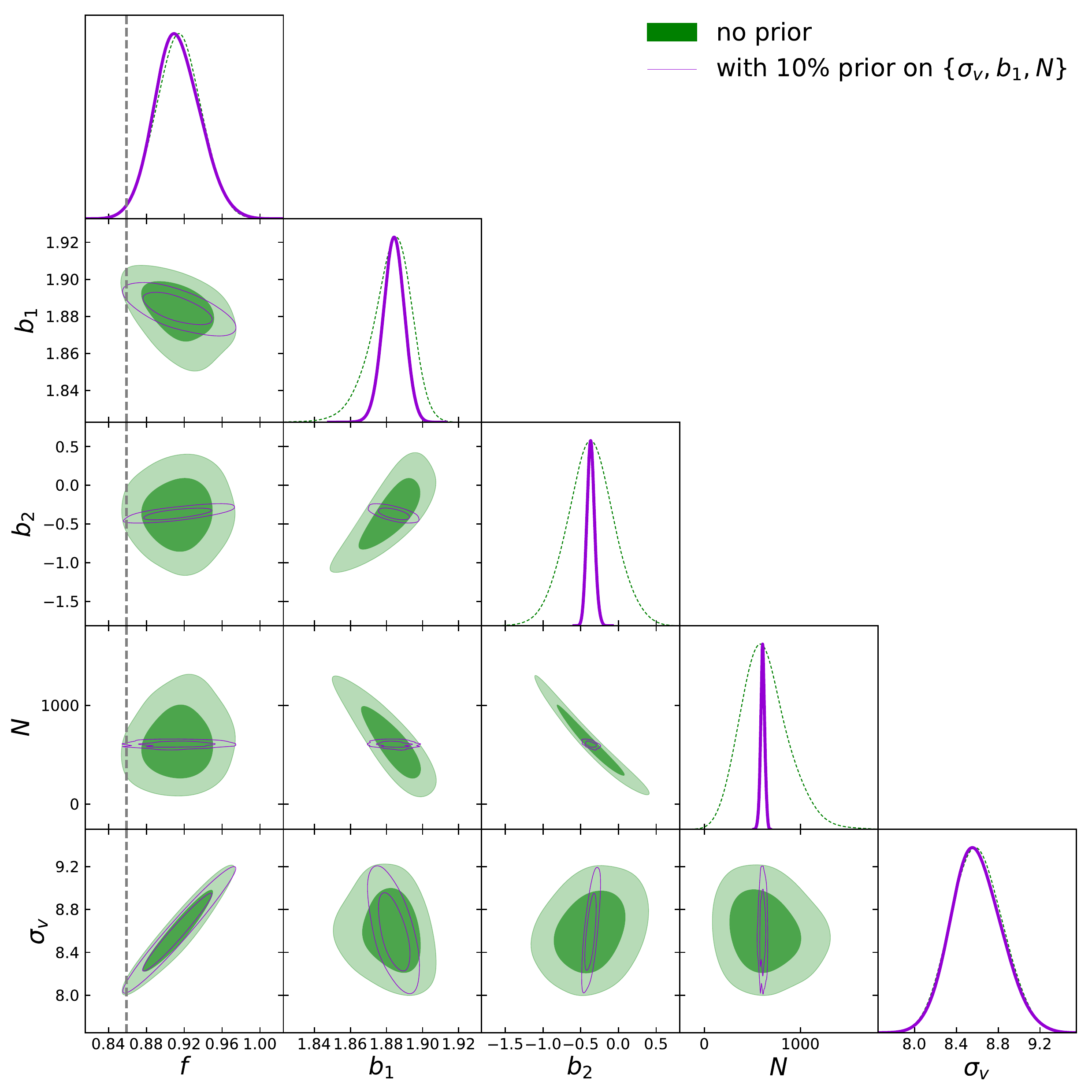}  
  \caption[]{MCMC results for the TNS model of \autoref{redshiftps} at $z=1$ with $k_{\rm max}~=~0.276~\,~h$/Mpc for $P_0$ and $P_2$ and $k_{\rm max,4}~=~0.05~\,~h$/Mpc for $P_4$. The dashed line indicates the fiducial value of $f=0.858$ in the PICOLA simulations. We also show the results with a $10\%$ prior around the best fit values of $\sigma_v,b_1$ and $N$. The best fit value of $f$ with $1 \sigma$ errors without (with) priors is $f=0.914\pm^{0.024}_{0.024}$ ($f=0.913\pm^{0.023}_{0.023}$).}
\label{mcmc3}
\end{figure*}

\begin{figure*}
\centering
  \includegraphics[height=0.8\textwidth,width=0.9\textwidth]{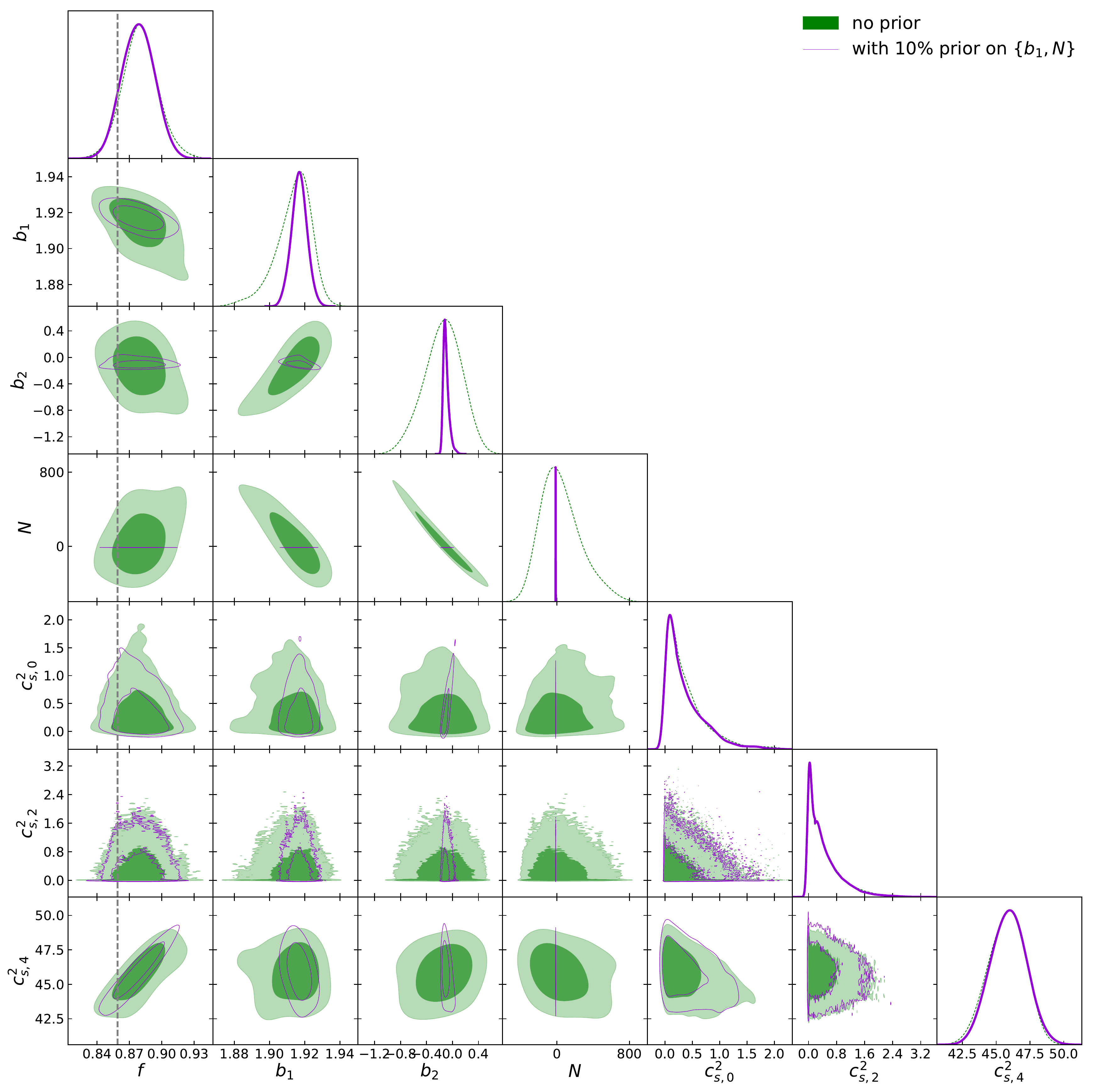}  
  \caption[]{MCMC results for the EFTofLSS model of \autoref{redshiftps2} at $z=1$ with $k_{\rm max}~=~0.276~\,~h$/Mpc for $P_0$ and $P_2$ and $k_{\rm max,4}~=~0.160~\,~h$/Mpc for $P_4$. The dashed line indicates the fiducial value of $f=0.858$ in the PICOLA simulations. We also show the results with a $10\%$ prior around the best fit values of $b_1$ and $N$. The best fit value of $f$ with $1 \sigma$ errors without (with) priors is $f=0.880\pm^{0.016}_{0.016}$ ($f=0.879\pm^{0.015}_{0.016}$).}
\label{mcmc4}
\end{figure*}

 In this section we present the results of a comprehensive MCMC analysis performed at $z=0.5$ and $z=1$. We use \autoref{covarianceeqn} to model our log-likelihood and vary the nuisance parameters outlined in \autoref{sec:models} as well as the growth rate $f$. We impose the same priors as when determining the minimum $\chi^2$ in \autoref{sec:sims-comparison}, i.e. $b_1,\sigma_v,c_{s,i}>0$, and use linear theory for the covariance matrix. This approach provides a more robust and accurate indication of each model's capability with respect to growth constraints, as well as parameter degeneracies. 
 \newline
 \newline 
 Furthermore, we will also consider the hexadecapole.  For the TNS model, it has been found that taking the hexadecapole up to the $k_{\rm max}$ shown in \autoref{fittable} produces  a biased estimate of the growth rate $f$. This is because the model is not flexible enough to account for the hexadecapole up to this high $k_{\rm max}$; note that this has been seen in the BOSS analysis \citep[e.g][]{Beutler:2016arn} as well as the TNS-Lorentzian forecast analysis in \citetalias{Markovic:2019sva}. Thus, to proceed we consider it up to a conservative value, $k_{\rm max,4}$, while taking the monopole and quadrupole up to the $k_{\rm max}$ found in \autoref{fittable}. Note that this is different from what was done in the Fisher analysis of \autoref{sec:forecasts}, which used the full $P(k,\mu)$ power spectrum up to a single $k_{\rm max}$ value. Instead, the procedure here closely resembles that followed in real data analyses, for example in \citet{Beutler:2016arn}.  For the EFTofLSS, we test different $k_{\rm max,4}$ for both $z=0.5$ and $z=1$. We find this prescription is capable of modelling the hexadecapole in an unbiased way up to a higher $k_{\rm max,4}$, an expected result because of its additional free parameters. 
 \newline
 \newline
 At $z=0.5$ we use $k_{\rm max,4} = 0.129 \, h/{\rm Mpc}$  for the TNS model, which is slightly larger than the value chosen at a similar redshift in \citet{Beutler:2016arn} ($k_{\rm max,4} = 0.100 \, h/{\rm Mpc}$), but we find this does not produce biased estimates for $f$ (while a larger value of $k_{\rm max,4}$ does).  For the EFTofLSS, we use $ k_{\rm max, 4} = k_{\rm max} = 0.245 \, h/{\rm Mpc}$. These results are shown for the TNS and EFTofLSS models in \autoref{mcmc1} and \autoref{mcmc2} respectively. In the same figure we plot contours that repeat the same analysis while also imposing $10\%$ flat priors on the best fit values of $\{b_1,N\}$ as well as $\sigma_v$ for TNS, similarly to what was done in \autoref{sec:forecasts}.
\newline
\newline
Next we consider $z=1$. Based on \citetalias{Markovic:2019sva} we take $k_{\rm max,4} = 0.05 \, h/{\rm Mpc}$ for the TNS model,   while for the EFTofLSS we find that taking $k_{\rm max,4} > 0.16 \, h/{\rm Mpc}$ biases the results. We plot these cases along with the same analyses using $10\%$ flat priors on the best fit values of $\{b_1,N\}$ (as well as $\sigma_v$ for TNS) in \autoref{mcmc3} and \autoref{mcmc4}. 
 The reader might wonder why $k_{\rm max,4}$ is lower at this redshift compared to the one at $z=0.5$. This can be explained through the $k_{\rm max}$ used for $P_0$ and $P_2$. At $z=1$, $k_{\rm max}$ is significantly higher than at $z=0.5$, hence the model's flexibility is being more severely tested. As a result, we find the models do not have the capacity to fit $P_4$ to a higher $k_{\rm max,4}$. To further elucidate this, we refer the reader to \autoref{redc} where we clearly see the reduced $\chi^2$ is lower and closer to $1$ at $z=0.5$ than at $z=1$ at the chosen $k_{\rm max}$.
\newline
\newline
We summarise all the marginalised $1\sigma$ percent errors on $f$ in \autoref{constraintstab} along with constraints coming from an analysis only using $P_0$ and $P_2$. 
We find that imposing the priors gives no significant change in the marginalised constraints of either model at either redshifts. In the  TNS and EFTofLSS cases at $z=0.5$, imposing the prior even worsens the constraint.  Taking the TNS case as an example, we find this to be a marginalisation effect related to the prior on $N$. The prior moves the entire posterior to larger values of $N$, which after marginalisation leads to larger errors on $f$ (see \autoref{mcmc3}). Changing the mean value of N to a smaller value (close to 0) before applying the $10\%$ prior, marginally reduces the percent error on $f$ from $3.2\%$ to $3.1\%$. 
\newline
\newline
In contrast to the exploratory, full $P(k,\mu)$ Fisher matrix analysis performed in \autoref{sec:forecasts}, at $z=0.5$ we find that the EFTofLSS model does significantly better than the TNS model and the gain from the inclusion of the hexadecapole in the EFTofLSS model is also larger with respect to the TNS case. At $z=1$, where the models have the same range of validity, the improvement is less dramatic. An important point we wish to reemphasise is that taking the hexadecapole up to too high a $k_{\rm max}$ (the ones found in \autoref{fittable}) produces biased estimates of the growth rate $f$ for both models,  with the exception of the EFTofLSS case at $z=0.5$ where indeed we can take $k_{\rm max}=k_{\rm max,4}$. This is what has been done in the Fisher analysis in \autoref{sec:forecasts}, which uses the full $P(k,\mu)$ up to the same $k_{\rm max}$ from \autoref{fittable}. As we have already stated, the MCMC analysis resembles what is done in a real data analysis procedure, and is therefore more robust and reliable. 
\newline
\newline
 To reiterate the point above, we have also performed checks to see what happens if we set the same $k_{\rm max}=k_{\rm max,4}$ for TNS at $z=0.5$. We found that including the hexadecapole at the same $k_{\rm max}$ results in $k_{\rm max}=k_{\rm max,4}=0.135 \, {\rm h/Mpc}$ for TNS giving a (barely) unbiased result for $f$. This results to a larger error of $4.5\%$, compared to our default case with $3.2\%$ error.  We have also checked that increasing$k_{\rm max}=k_{\rm max,4}$ beyond $0.135 \, {\rm h/Mpc}$ results in a biased recovery of $f$. A similar test can be found in Figure 4 of \citetalias{Markovic:2019sva}. 
Further, we note that in earlier work by \citet{Taruya:2011tz}, Fisher matrix forecasts using the multipole expansion were performed for the TNS model equipped with a linear bias prescription, taking the same $k_{\rm max}$ for monopole, quadrupole, and hexadecapole. The authors showed that the monopole and quadrupole contain most of the constraining power, but the hexadecapole can somewhat help to further decrease the errors. While the model considered in \citet{Taruya:2011tz} is not the same as the one we consider in this work, the main result is general: it is preferable to include the monopole and quadrupole at a derived common $k_{\rm max}$ and then add the hexadecapole for as long as the constraints are not biased.

\subsection{The effect of positivity priors on the EFTofLSS constraints}

It is important to comment on the effect of the positivity priors imposed on the EFTofLSS parameters $c^2_{s,i}$; these can have a non-negligible effect on the EFTofLSS $f$ constraints. More specifically, we have found that if the fiducial values for these parameters are sufficiently away from zero so that no positivity priors are needed, the $f$ constraints can increase substantially. This explains the apparent inconsistency between our MCMC results and some of the MCMC results in \citetalias{Bose:2019ywu}. In this paper a different set of simulations, cosmology, and method to determine model ranges validity were used. Specifically, for the halo catalog they consider the authors of \citetalias{Bose:2019ywu} find that the best fit values for $c_{s,i}^2$ are high enough so as not to run into the positivity priors imposed on these nuisance parameters. This results in worse marginalised constraints on $f$. We have studied the difference this makes by comparing with \citetalias{Bose:2019ywu} Figure 12 as an example, and we have found that indeed the Fisher and MCMC results agree perfectly with no $c^2_{s,i}$ priors needed. This explains the fact that the TNS model is found to outperform EFTofLSS in the analysis of \citetalias{Bose:2019ywu}. This is an important subtlety as these parameters can have very different values for different galaxy samples or different theories of gravity and dark energy. This suggests that validation studies should be performed on a case to case basis, and that fast and reliable Fisher forecasts like the ones performed here can be particularly helpful at the initial validation stages.

\begin{table*}
\centering
\caption{$1\sigma$ marginalised percent errors on $f$ from the MCMC analyses at $z=0.5$ and $z=1$. We utilise the monopole and quadrupole up to the $k_{\rm max}$ given in \autoref{fittable},  and the hexadecapole up to $k_{\rm max,4} = 0.129 \, h/{\rm Mpc}$ for TNS and $k_{\rm max,4} = 0.245 \, h/{\rm Mpc}$ for EFTofLSS at $z=0.5$, while $k_{\rm max,4} =0.05 \, h/{\rm Mpc}$ for TNS and $k_{\rm max,4} =0.16 \, h/{\rm Mpc}$ for EFTofLSS at $z=1$.}
\begin{tabular}{| c || c | c | c | c | c | c |}
\hline  
 \multicolumn{1}{ | c || }{} & \multicolumn{3}{|c|}{TNS Lor} &  \multicolumn{3}{|c|}{EFTofLSS} \\
 \hline
  \multicolumn{1}{ | c || }{$z$} & $P_0+P_2$ &  $+P_4$ &  $+10\%$ prior
  & $P_0+P_2$ &  $+P_4$ & $+10\%$ prior \\
 \hline\hline
0.5 & $3.6\%$ &  $3.2\%$ &  $ 3.5 \%$ &   $2.8\%$  &  $1.8\%$  &  $2.1\%$ \\ \hline 
1 & $3.0\%$ &  $2.6\%$ &  $ 2.5 \%$ &   $2.0\%$  &   $1.8\%$  &  $1.7\%$ \\ \hline 
\end{tabular}
\label{constraintstab}
\end{table*}



\section{MCMC and Fisher matrix comparison for EFTofLSS using multipole expansion}
\label{sec:FishMCMCcomp}
\begin{figure*}
\centering
  \includegraphics[height=0.6\textwidth,width=0.7\textwidth]{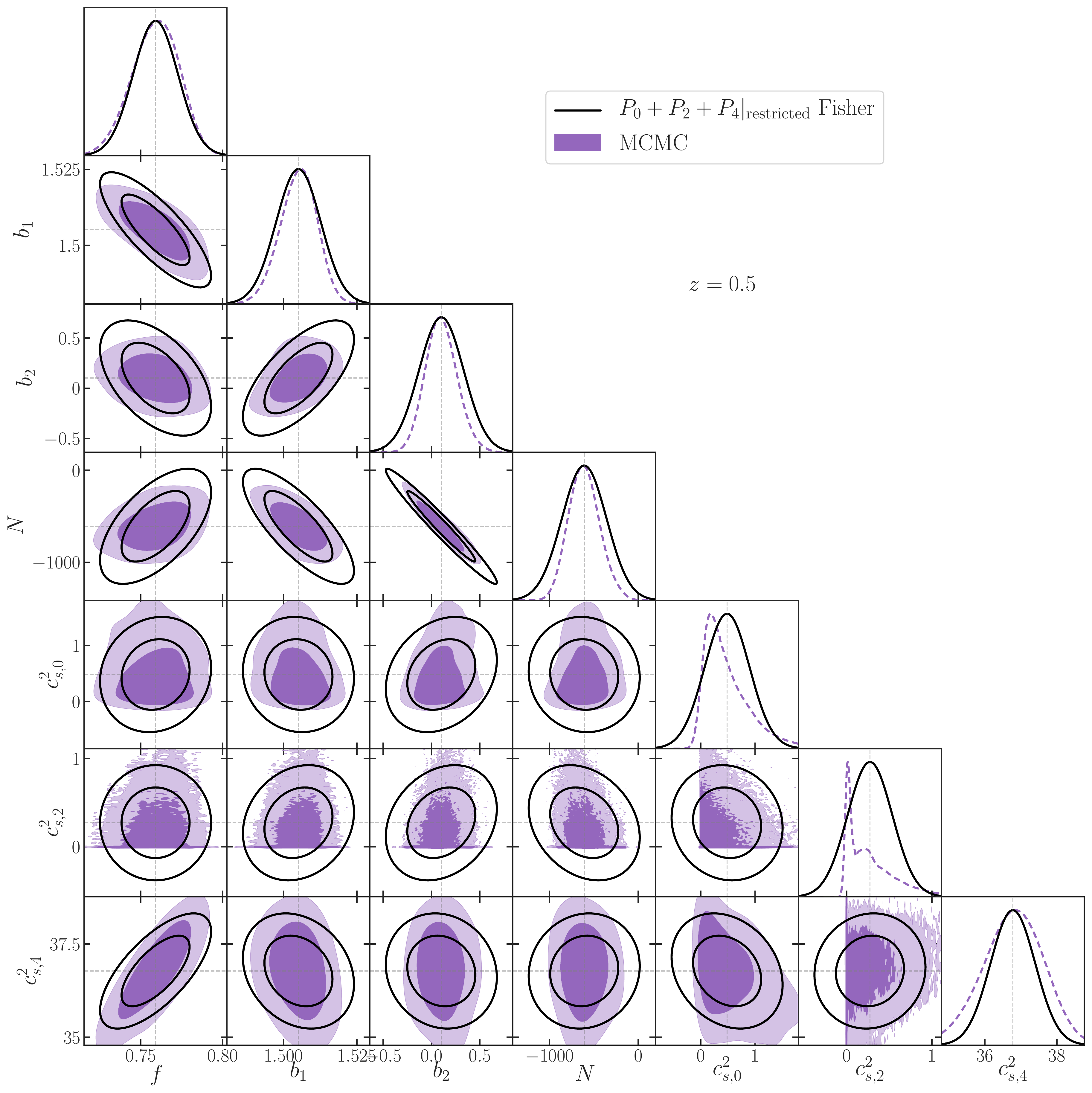}  
  \caption[]{Fisher matrix and MCMC comparison for the EFTofLSS model of \autoref{redshiftps2} at $z=0.5$, for the $P_0+P_2+P_4|_{\rm restricted}$ case.}
\label{compzeq0p5}
\end{figure*}

\begin{figure*}
\centering
  \includegraphics[height=0.6\textwidth,width=0.7\textwidth]{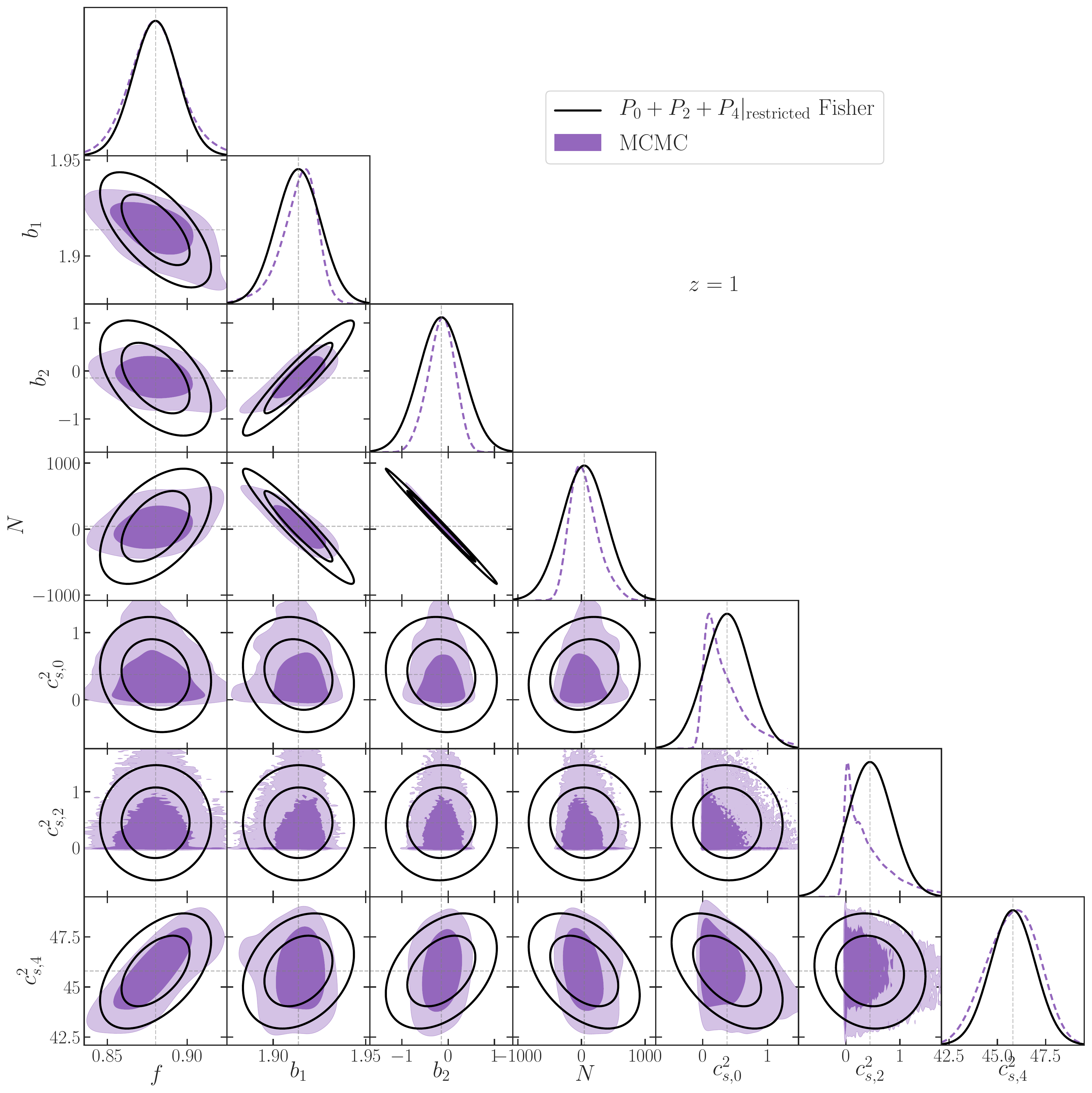}  
  \caption[]{Fisher matrix and MCMC comparison for the EFTofLSS model of \autoref{redshiftps2} at $z=1.0$, for the $P_0+P_2+P_4|_{\rm restricted}$ case.}
\label{compzeq1}
\end{figure*}

\begin{figure*}
\centering
  \includegraphics[height=0.6\textwidth,width=0.7\textwidth]{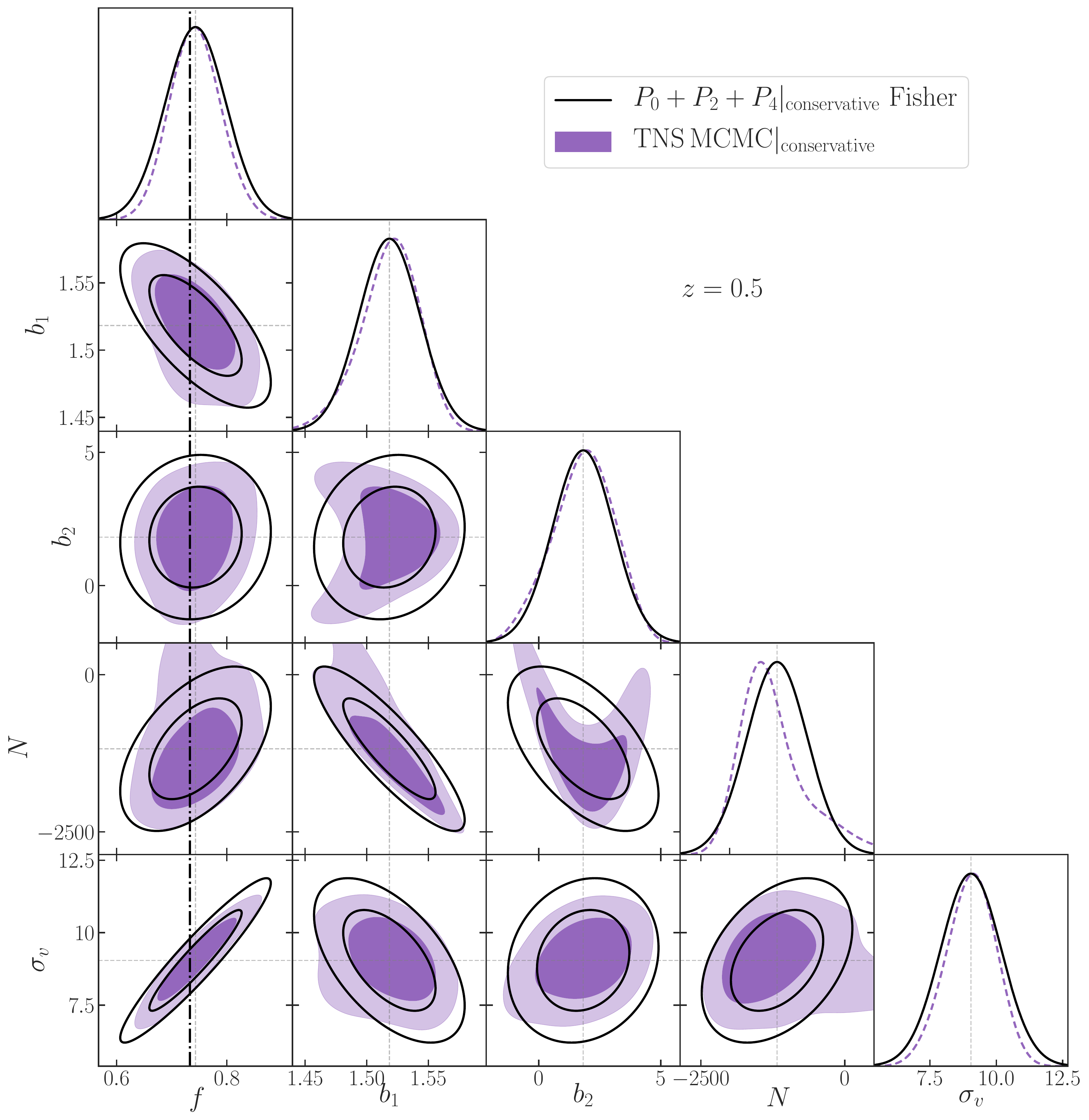}  
  \caption[]{Fisher matrix and MCMC comparison for the TNS model at $z=0.5$, for the $P_0+P_2+P_4|_{\rm conservative}$ case. The dashed line indicates the fiducial value of $f=0.733$ in the PICOLA simulations. The best fit value of $f$ with $1 \sigma$ errors is $f=0.743\pm^{0.046}_{0.046}$.}
\label{compzeq0p5_TNS_cons}
\end{figure*}

\begin{figure*}
\centering
  \includegraphics[height=0.6\textwidth,width=0.7\textwidth]{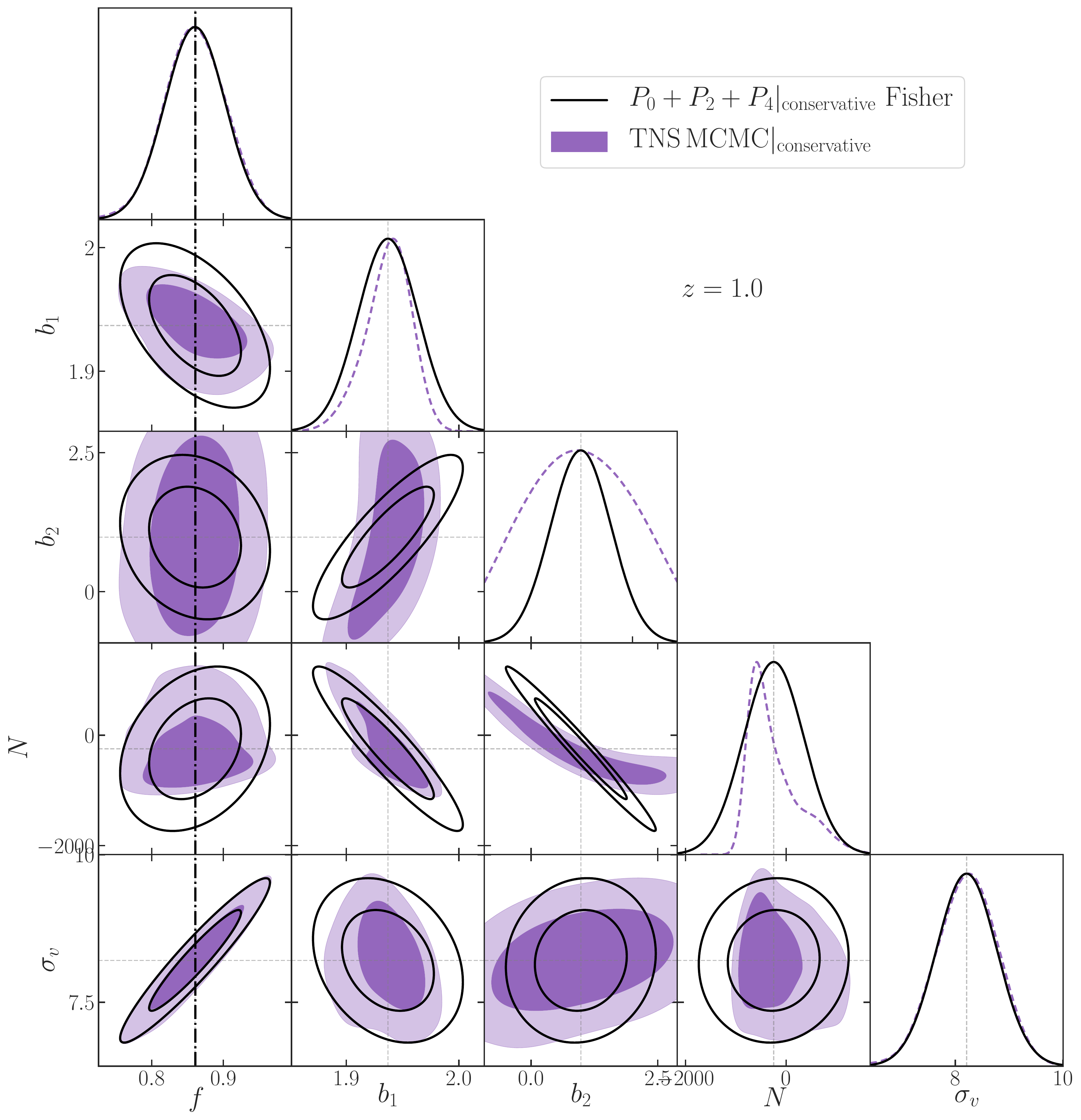} 
  \caption[]{Fisher matrix and MCMC comparison for the TNS model at $z=1$, for the $P_0+P_2+P_4|_{\rm conservative}$ case. The dashed line indicates the fiducial value of $f=0.858$ in the PICOLA simulations. The best fit value of $f$ with $1 \sigma$ errors is $f=0.861\pm^{0.042}_{0.042}$.}
\label{compzeq1_TNS_cons}
\end{figure*}

\begin{figure*}
\centering
  \includegraphics[height=0.6\textwidth,width=0.7\textwidth]{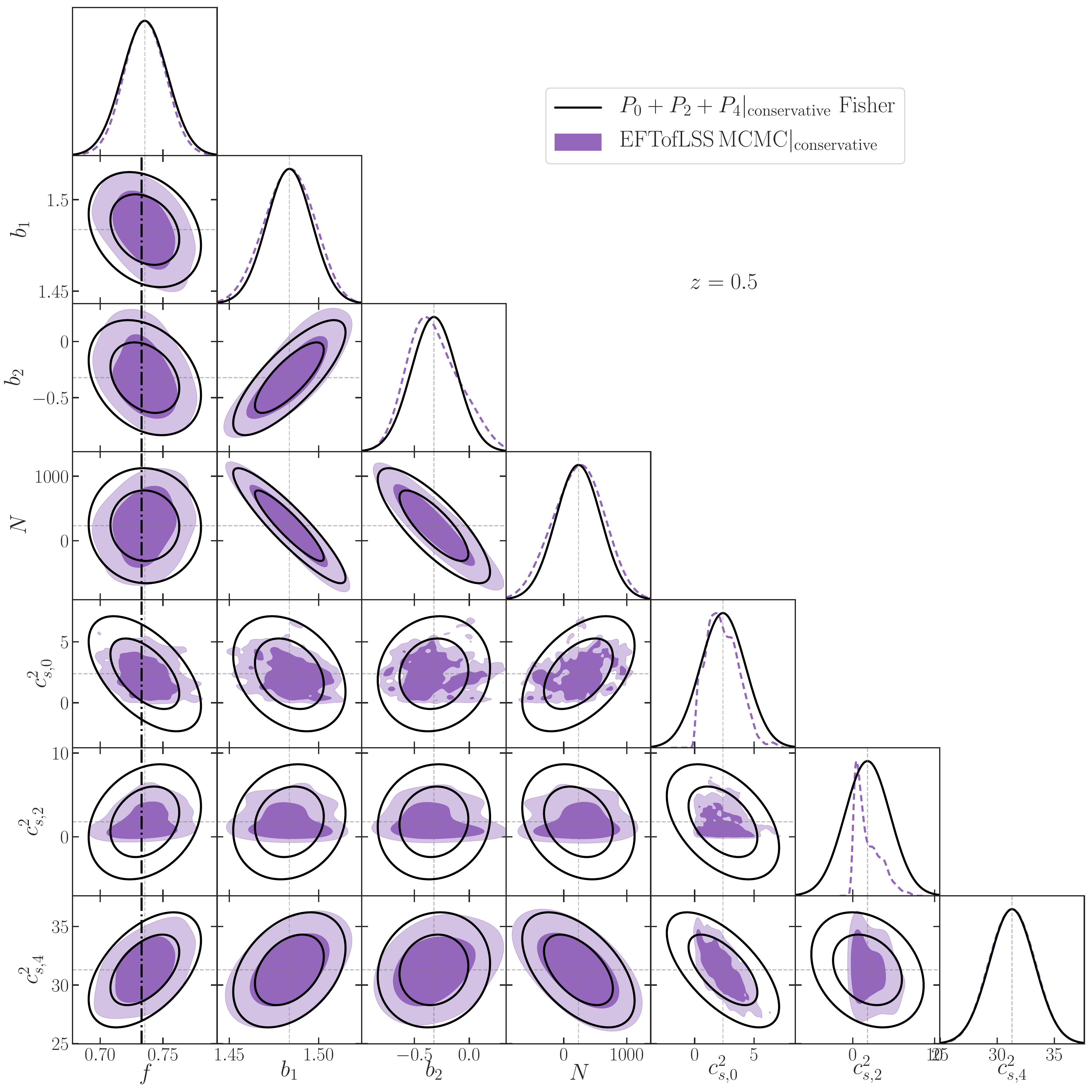}  
  \caption[]{Fisher matrix and MCMC comparison for the EFTofLSS model at $z=0.5$, for the $P_0+P_2+P_4|_{\rm conservative}$ case. The dashed line indicates the fiducial value of $f=0.733$ in the PICOLA simulations. The best fit value of $f$ with $1 \sigma$ errors is $f=0.736\pm^{0.016}_{0.016}$.}
\label{compzeq0p5_EFT_cons}
\end{figure*}

\begin{figure*}
\centering
  \includegraphics[height=0.6\textwidth,width=0.7\textwidth]{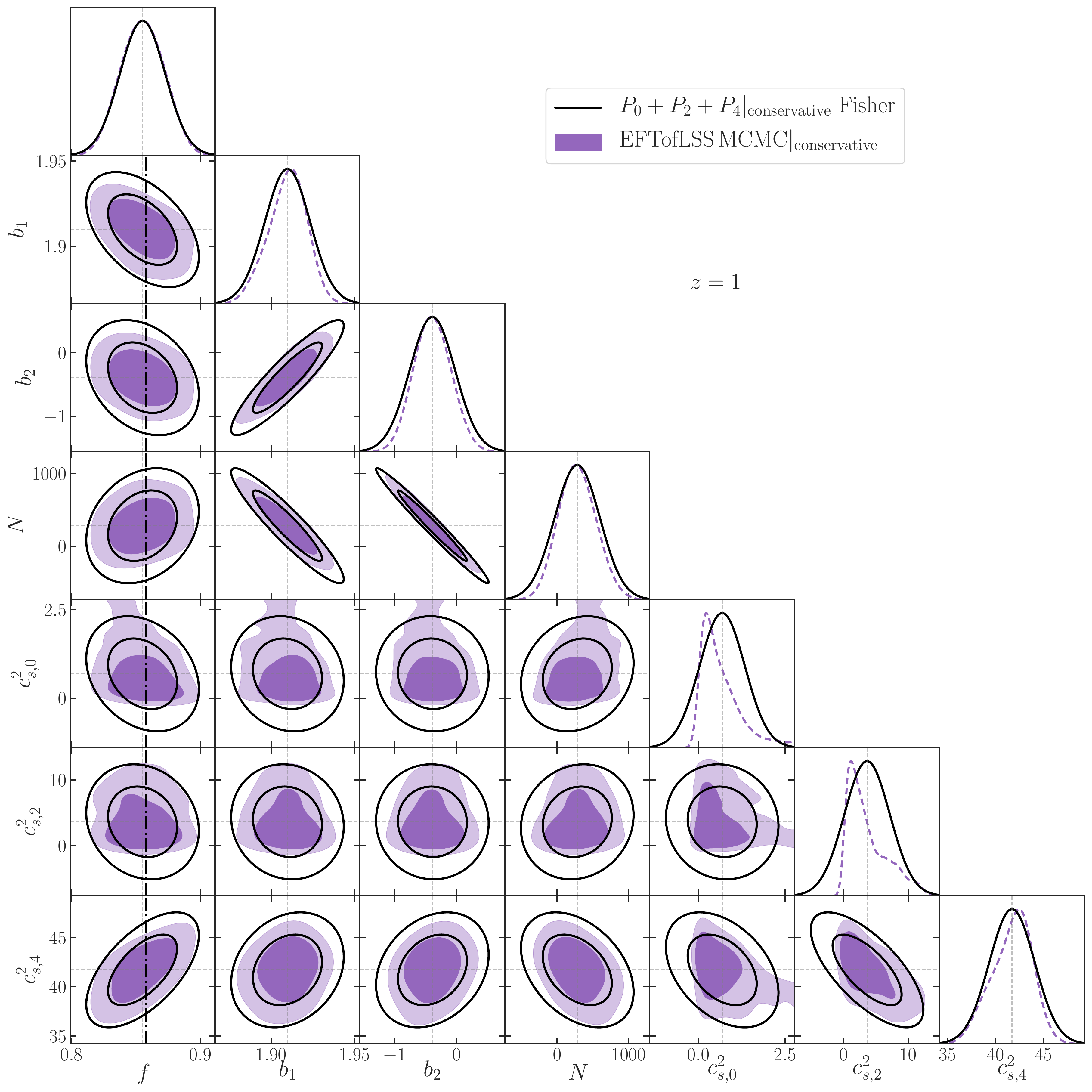}  
  \caption[]{Fisher matrix and MCMC comparison for the EFTofLSS model at $z=1$, for the $P_0+P_2+P_4|_{\rm conservative}$ case. The dashed line indicates the fiducial value of $f=0.858$ in the PICOLA simulations. The best fit value of $f$ with $1 \sigma$ errors is $f=0.855\pm^{0.017}_{0.017}$.}
\label{compzeq1_EFT_cons}
\end{figure*}

Having calculated both, the Gaussian approximation to the likelihood using the Fisher formalism as well as the full non-Gaussian likelihood using the MCMC technique, we would like to assure ourselves that they give concordant results, allowing for some discrepancies from approximating. However, it has been shown in the TNS-Lorentzian case \citepalias{Markovic:2019sva} that the high-$k$ contribution of the hexadecapole can give deceptively good error predictions, when using the full, 2-dimensional $P(k,\mu)$ as the observable in the Fisher matrix. So, instead, we now consider the Fisher matrix of the power spectrum multipoles in \autoref{eq:ell} in order to be able to exclude that contribution, that in the real analysis, would result in a biased best estimate of our cosmological parameter, $f$. The multipole Fisher matrix is described in \citet{Taruya:2011tz} and \citetalias{Markovic:2019sva} (the latter using precisely the same conventions as this paper). As in \citetalias{Markovic:2019sva}, we call this Fisher multipole analysis $P_0+P_2+P_4|_{\rm restricted}$.
\newline
\newline
We do this only for the EFTofLSS model here and refer the reader to \citetalias{Markovic:2019sva} for the analysis using the TNS-Lorentzian case. As before, we only consider the Gaussian covariance between the observables, which means that the covariance between different $k$-modes is approximated to be zero. Furthermore, we now use the means of the MCMC analysis as the fiducial values of the Fisher matrix multipole analysis, to allow for a consistent comparison.
\newline
\newline
In \autoref{compzeq0p5} and \autoref{compzeq1} we show the resulting posteriors at $z=0.5$ and $z=1.0$ respectively, shaded for the MCMC and lines showing the Gaussian Fisher matrix contours. We find very good agreement in our cosmological parameter $f$ between the two approaches, but note discrepancies in the EFTofLSS nuisance parameters. This means that the discrepancy propagates only minimally into the marginalised posterior for $f$. These discrepancies between the MCMC and Fisher of the nuisance parameters may be a result of asymmetric true posteriors for the $c^2_{s,i}$ parameters, which cannot take negative values. This feature is not visible to the Fisher matrix, since it can only ever describe Gaussian likelihoods.
In order to mitigate this issue, we include conservative Gaussian priors on our EFTofLSS nuisance parameters, $c^2_{s,i}$, with their $\sigma \sim 100\%$ fiducial value. This cannot exclude the negative region for the $c^2_{s,i}$ parameters, but it can help make it less likely.
As in \citetalias{Markovic:2019sva}, we also notice some very large correlation coefficients between $b_2$ and $N$. Such correlations can induce instabilities in the inversion of the Fisher matrix (needed to calculate the parameter covariance), so we impose a conservative prior on $N$ as well.
 Investigating such priors in more detail would be worthwhile, but would require running suites of MCMC to validate against. The marginalised $1\sigma$ constraints from these analyses can be found in \autoref{summarytable} and, as in the TNS case \citepalias{Markovic:2019sva}, are very consistent with the MCMC results. 
\newline
\newline
 Finally, to consolidate our findings we have also performed a comparison between \emph{conservative} Fisher matrix and MCMC forecasts. For the TNS model at $z=0.5$ we consider $k_{\rm max}=k_{\rm max,4} = 0.1 \, h/{\rm Mpc}$ and the results are shown in \autoref{compzeq0p5_TNS_cons}. At $z=1$ we take $k_{\rm max} = 0.15 \, h/{\rm Mpc}$ and $k_{\rm max,4} = 0.05 \, h/{\rm Mpc}$, and the results are shown in \autoref{compzeq1_TNS_cons}. In both cases the $f$ parameter estimation is unbiased at a level smaller than $1\sigma$. The fractional MCMC $f$ errors are $6.1\%$ at $z=0.5$ and $4.8\%$ at $z=1$.  For the EFTofLSS model at $z=0.5$ we take $k_{\rm max}=0.245 \, h/{\rm Mpc}$, $k_{\rm max,4}=0.129 \, h/{\rm Mpc}$, and the results are shown in \autoref{compzeq0p5_EFT_cons}. At $z=1$ we take $k_{\rm max} = 0.276 \, h/{\rm Mpc}$ and $k_{\rm max,4} = 0.08 \, h/{\rm Mpc}$, and the results are shown in \autoref{compzeq1_EFT_cons}. In both cases the $f$ parameter estimation is unbiased at a level smaller than $1\sigma$. The fractional MCMC $f$ errors are $2.2\%$ at $z=0.5$ and $2.0\%$ at $z=1$. We note that these results are not to be used as a means of comparison between the constraining power of the models -- the $k_{\rm max}$ chosen are very different and they do not correspond to the maximum $k_{\rm max}$ for the $f$ estimation to be unbiased at less than $1\sigma$; they were chosen empirically.  As expected from our previous results, in all cases we find very good agreement in the $f$ Fisher matrix constraints despite differences in the nuisance parameters due to the non gaussian shape of several of the ellipses.


\section{Summary and Conclusions}
\label{sec:conclusions}

This paper is the second in a series of three. \citetalias{Markovic:2019sva} demonstrated how to calculate Fisher matrix forecasts for Stage IV galaxy surveys using the multipole expansion of the redshift-space galaxy power spectrum. It validated the Fisher matrix method with multipoles against a full MCMC analysis, all using the TNS model. It also demonstrated that not accounting for inaccurate modelling of the small scales of the hexadecapole not only biases the location of the maximum likelihood (as shown by previous works), it also results in an overly optimistic forecast.
\newline
\newline
In this paper we have built on the conclusions of \citetalias{Markovic:2019sva} and compared two prominent models for the redshift space halo power spectrum in the context of upcoming galaxy surveys: the commonly used TNS model and an EFTofLSS-based model, equipped with $4$ and $6$ nuisance parameters respectively. These models are very similar to the  \texttt{M\&R+EFT} and \texttt{M\&R+SPT} models considered in \citet{delaBella:2018fdb}. The EFTofLSS-based model presented here has a largely reduced nuisance parameter set than the full biased tracer model of \citet{Perko:2016puo} (10 nuisance parameters) and that considered in \citet{delaBella:2018fdb} (8 nuisance parameters). We consider two redshifts, $z=0.5$ and $z=1$ and make use of 4 realisations of $V=1 \, \mbox{Gpc}^3/h^3$ PICOLA simulations to perform maximum likelihood, Fisher matrix, and MCMC analyses. Here we summarise our main results and conclude. All core results are presented in \autoref{summarytable}.
\\ \\
{\bf Model ranges of validity}: 
We determine ranges of validity by imposing the best fit $\chi_{\rm red}^2 \lessapprox 1$ using only the monopole and quadrupole. Errors are determined using linear theory and specifications similar to a Stage IV spectroscopic galaxy survey: a survey (bin) volume $V_s=4 \, \mbox{Gpc}^3/h^3$ and a tracer number density $n=1\times10^{-3} \, h/\mbox{Mpc}$. Our results are: 
\begin{enumerate}
    \item 
The TNS model equipped with a Lorentzian damping factor (TNS-Lorentzian) greatly out-performs the same model equipped with a Gaussian damping factor at both $z=0.5$ and $z=1$. 
\item
The TNS-Lorentzian model employing an SPT prescription for the 1-loop spectrum terms out-performs the same model using a RegPT prescription (as used in the BOSS analysis) at both $z=0.5$ and $z=1$.
\item 
The TNS-Lorentzian performs similarly to the EFTofLSS model at $z=1$ with a shared $k_{\rm max} = 0.276\, h/{\rm{Mpc}}$. At $z=0.5$ the EFTofLSS model does well up to $k_{\rm max}=0.245 \, h/{\rm{Mpc}}$ while the TNS up to a lower $k_{\rm max}=0.227 \, h/{\rm{Mpc}}$. This is attributed to the EFTofLSS's ability to model the enhanced non-linearity at lower redshift using its additional nuisance parameters.
\end{enumerate}
{\bf Fisher analysis using the full $P(k,\mu$)}: 
We perform an exploratory Fisher analysis on the TNS-Lorentzian and EFTofLSS-based models using the ranges of validity found in \autoref{sec:sims-comparison} and the full $P(k,\mu)$. In addition to the nuisance parameters we also vary the logarithmic growth rate, $f$. Our results are summarised in \autoref{Fisherconstraintstab}.
The analysis using the $k_{\rm max}$ from \autoref{fittable} shows a significant degeneracy between $f$ and $\sigma_v$ for the TNS model which has also been found previously \citep{Zheng:2016xvo,Bose:2017myh}. The improvement on the TNS constraints at $z=1$ is mainly due to the much higher $k_{\rm max}$ at $z=1$ compared to that at $z=0.5$. 
\\ 
\\
For the EFTofLSS-based model, the constraints are practically the same for the two redshifts (slightly better at $z=1$), and worse than the constraints using the TNS model at both redshifts. At $z=0.5$, where non-linear effects are more important at lower $k$, we see that the EFTofLSS-based model allows us to use a larger $k_{\rm max}$ than the TNS model but the final, marginalised $f$ constraints are better with TNS. At $z=1$ the two models have the same $k_{\rm max}$, but the degeneracies between nuisance parameters and $f$ result to a better constraint using TNS.  This conclusion holds at $z=1$ even when considering the same conservative $k_{\rm max}=0.15h/{\rm Mpc}$ for both models.
\\
\\
Knowing that this analysis is just exploratory (mainly due to the restricted range of scales required for the hexadecapole in order for the $f$ estimation to be unbiased - see  \citetalias{Markovic:2019sva}), we moved on to an MCMC analysis.
\\\\
\noindent
{\bf MCMC analysis}: 
We perform two distinct MCMC analyses at $z=0.5$ and $z=1$ on the TNS-Lorentzian and EFTofLSS models including the first 3 multipoles of the RSD power spectrum, $P_0$, $P_2$ and $P_4$, and in which we vary all nuisance parameters along with the logarithmic growth rate of structure $f$. For $P_0$ and $P_2$ we use the range of validity, $k_{\rm max}$, determined in \autoref{sec:sims-comparison}, while for the hexadecapole we restrict its range to a lower $k_{\rm max,4}$ that is checked not to bias the estimation of $f$ within $2\sigma$, similar to what was done in the BOSS analysis of \citet{Beutler:2016arn}. 
In all MCMC analyses the fiducial growth rate is recovered within the $2\sigma$ region. Our main results are: 

\begin{enumerate}
    \item 
    At $z=0.5$, the inclusion of the hexadecapole noticeably improves the marginalised $1\sigma$ constraints on $f$. For the TNS model we get $\sim 12\%$ improvement by including $P_4$ while for EFTofLSS we get $\sim 36\%$ improvement in the constraints without considering any priors.
    \item 
    At $z=1$ without any priors, the inclusion of the hexadecapole noticeably improves the marginalised $1\sigma$ constraints on $f$ again by $\sim 14\%$ for the TNS model. For the EFTofLSS model,  the constraints improve by $\sim 10\%$.
    \item
    The TNS model without priors (with a $10\%$ prior on $\{b_1,N,\sigma_v\}$) gives a $3.2\% \, (3.5\%)$ marginalised $1\sigma$ error on $f$ at $z=0.5$ and $2.6\% \, (2.5\%)$ error at $z=1$. 
    \item
     The EFTofLSS model without priors (with a $10\%$ prior on $\{b_1,N\}$) gives a $1.8\% \, (2.1\%)$ marginalised $1\sigma$ error on $f$ at $z=0.5$ and $1.8\% \, (1.7\%)$ error at $z=1$.
    \end{enumerate}
\noindent
 This analysis maps the posterior distributions and does not assume their Gaussianity as is required for Fisher matrix analysis. We find it to be more representative of a real data analysis procedure and thus more reliable in informing future surveys. 
\\\\
\noindent
{\bf Comparing MCMC results to the Fisher analysis using multipoles, $P^{(S)}_l$, for EFTofLSS}:
In order to be able to compare the results of our MCMC to the approximate Gaussian posteriors calculated from Fisher matrices, we perform another Fisher matrix analysis, this time using multipoles as our observable.  We do this in order to better emulate the real data analysis procedure, which excludes the high-$k$ regime to avoid biased estimates of cosmological parameters. This is only done for the EFTofLSS model with the TNS analysis having already been performed in \citetalias{Markovic:2019sva}. We show that in order for the two posteriors to be consistent, conservative priors must be applied to the Fisher matrix. This is to avoid the pitfalls of highly degenerate parameters as well as to account for the asymmetry of the likelihoods for the EFTofLSS sound speed parameters, which cannot be negative. Doing this, we find very consistent marginalised constraints on $f$ when compared to the MCMC analyses performed here, with and without priors on the selected nuisance parameters, as can be seen in \autoref{summarytable}. 
\newline
\newline
{\bf Outlook:} In this paper, the EFTofLSS model seems to outperform the TNS model in terms of its marginalised constraints on $f$ when we consider the MCMC analysis as our benchmark.   This is the opposite of the conclusions found in \citetalias{Bose:2019ywu}. We have determined that this discrepancy comes from the impact of the positivity priors for the $c_{s,i}$ parameters in our EFTofLSS-like model. In \citetalias{Bose:2019ywu}, the simulation measurements and cosmology are such that these have a lesser impact, producing worse marginalised constraints on $f$. In this paper however, the positivity priors play a significant role by restricting the parameter space in advance. On the other hand, our results are similar to what was found in the reduced $\chi^2$ analysis of \citet{delaBella:2018fdb}, albeit with slightly different models to those used here (and different survey assumptions), with the closest being their \texttt{M\&R+EFT} and \texttt{M\&R+SPT} models. In that work they consider a redshift of $z=0.44$ and use more simulation realisations than here. Furthermore, they fit the hexadecapole all the way up to their fixed $k_{\rm max}$ of $0.290 \, h/{\rm Mpc}$. Interestingly, they find the EFTofLSS model is disfavoured if one considers a Bayesian information criterion to penalise each model depending on its number of nuisance parameters.
\newline
\newline
Despite having performed a number of complimentary analyses in this work, our investigation is far from exhaustive. For example, our determination of $k_{\rm max}$ does not vary $f$ and degeneracies between this and nuisance parameters may allow validity of the models to larger $k$. This requires broader MCMC analyses. This will be important to truly determine if there is a favoured model.  This being said, we have shown that the multipole decomposition within a Fisher framework can provide reliable and robust tests for the RSD models considered here, both in terms of parameter degenerecies and more so in terms of their cosmological constraining power. This suggests that the Fisher method can prove to be a reliable and extremely powerful tool in model selection for future surveys. 
\\
\\
 In \citetalias{Bose:2019ywu} we present an MCMC analysis studying the TNS-Lorentzian and EFTofLSS models presented here, using a larger suite of simulations to examine different $k_{\max}$, redshifts, survey volumes, and halo mass cuts which will aim at further separating the two competing models.
\\
\\
In future work it will also be interesting to consider the effect of including the bispectrum, as it should provide useful additional information \citep[see, for example,][]{Yankelevich:2018uaz}. We leave this for future work.
\begin{table*}
\centering
\caption{{\bf Summary of Results:} $1\sigma$ marginalised percent errors on $f$ from multipole expansion analyses performed in this work. The $k_{\rm max}$ used for $P_0$ and $P_2$ can be found in \autoref{fittable}.  For $z=0.5$ TNS uses the hexadecapole up to $k_{\rm max,4} = 0.129 \, h/{\rm Mpc}$ while in the EFTofLSS we have $k_{\rm max,4} = k_{\rm max} = 0.245 \, h/{\rm Mpc}$.
For $z=1$, $k_{\rm max,4} =0.05 \, h/{\rm Mpc}$ for TNS and $k_{\rm max,4} =0.16 \, h/{\rm Mpc}$ for EFTofLSS.} Bracketed quantities indicate the result using a $10\%$ prior applied on the parameter set $\{b_1,N\}$ as well as $\sigma_v$ for TNS. The Fisher:  $P_0+P_2+P_4|_{\protect\mathrm{restricted}}$ TNS case has been included here for completeness, but was calculated in \citetalias{Markovic:2019sva}. 
\begin{tabular}{| c | c | c | c | c |}
\hline  
\multicolumn{1}{ | c | }{} & \multicolumn{2}{|c|}{TNS Lor} &  \multicolumn{2}{|c|}{EFTofLSS} \\
 \hline
 Analysis & $z=0.5$ & $z=1$ & $z=0.5$ & $z=1$  \\
 \hline  \hline
 MCMC: $P_0+P_2$ & $3.6\%$ & $3.0\%$ & $2.8\%$  & $2.0\%$   \\ \hline 
 MCMC: $P_0+P_2+P_4|_{\mathrm{restricted}}$ & $3.2(3.5)\%$ & $2.6(2.5)\%$  &  $1.8(2.1)\%$ &  $1.8(1.7)\%$   \\ \hline 
 Fisher: $P_0+P_2+P_4|_{\mathrm{restricted}}$ & $3.8(3.5)\%$ & $2.9(2.8)\%$ &  $1.8(1.8)\%$ &  $1.6(1.4)\%$   \\ 
  \hline
\end{tabular}
\label{summarytable}
\end{table*}

\section*{Acknowledgments}
\noindent We are indebted to the anonymous referees whose suggestions greatly improved the quality of this manuscript. We are grateful to Hans Winther for providing the simulation data. We thank Lucia Fonseca de la Bella and David Seery for useful discussions. We acknowledge use of open source software \citep{scipy:2001,Hunter:2007, Lewis:1999bs, mckinney-proc-scipy-2010, numpy:2011}. BB acknowledges support from the Swiss National Science Foundation (SNSF) Professorship grant No.170547. AP is a UK Research and Innovation Future Leaders Fellow, grant MR/S016066/1, and also acknowledges support from the UK Science \& Technology Facilities Council through grant ST/S000437/1. KM acknowledges support from the UK Science \& Technology Facilities Council through grant ST/N000668/1 and from the UK Space Agency through grant ST/N00180X/1. KM's contribution was also partially carried out at the Jet Propulsion Laboratory, California Institute of Technology, under a contract with the National Aeronautics and Space Administration (80NM0018D0004). FB is a Royal Society University Research Fellow. Numerical computations for this research were done on the Sciama High Performance Compute (HPC) cluster which is supported
by the ICG, SEPNet, and the University of Portsmouth.



\bibliographystyle{mnras}
\bibliography{mybib}

\bsp	
\label{lastpage}
\end{document}